\pdfoutput=1
\documentclass[10pt,twocolumn,a4paper,twoside]{IEEEtran}
\usepackage[T1]{fontenc}
\usepackage[utf8]{inputenc}

\usepackage{csquotes}

\ifCLASSOPTIONdraftcls
\newcommand{\aligniftwocolumn}{}
\newcommand{\breakiftwocolumn}{}

\newcommand{\raiseintwocolumn}[1]{}
\newcommand{\breakiftwocolumnquad}{\quad}
\newcommand{\cdotiftwocolumn}{}
\else
\newcommand{\raiseintwocolumn}[1]{\raisetag{#1}}
\newcommand{\aligniftwocolumn}{&}
\newcommand{\breakiftwocolumn}{\\}

\newcommand{\breakiftwocolumnquad}{\\}
\newcommand{\cdotiftwocolumn}{\cdot}
\fi

\usepackage{amsmath}
\usepackage{amssymb}
\DeclareMathOperator{\diag}{diag}
\DeclareMathOperator{\E}{E}

\usepackage{mathtools}

\DeclarePairedDelimiter{\abs}{\lvert}{\rvert}
\DeclarePairedDelimiter{\norm}{\lVert}{\rVert}
\DeclarePairedDelimiter{\parens}{\lparen}{\rparen}
\DeclarePairedDelimiter{\bracks}{\lbrack}{\rbrack}
\DeclareMathOperator{\Real}{Re}

\DeclareMathOperator{\tr}{tr}

\usepackage{algpseudocode}

\usepackage{array}

\usepackage{url}
\usepackage[expansion]{microtype}
\input{glyphtounicode.tex}
\input{glyphtounicode-cmr.tex}
\usepackage{hyperref}
\hypersetup{pdfauthor={Laas, Tobias; Nossek, Josef A.; Bazzi, Samer; Xu, Wen},pdftitle={On Reciprocity in Physically Consistent TDD Systems with Coupled Antennas}}
\usepackage{doi}

\usepackage{bm}
\newcommand{\matt}[1]{\bm{#1}}
\newcommand{\vect}[1]{\bm{#1}}

\usepackage{siunitx}

\usepackage{etoolbox}
\newbool{finalsubmission}
\setbool{finalsubmission}{true}
\newbool{pdffigures}
\setbool{pdffigures}{false}
\ifbool{finalsubmission}{\setbool{pdffigures}{true}}{}
\ifbool{CLASSOPTIONdraftcls}{\setbool{pdffigures}{true}}{}

\usepackage{tikz}
\ifthenelse{\boolean{finalsubmission}}{}{
\usepackage[europeanresistors,straightvoltages]{circuitikz}
\usetikzlibrary{scopes}
\ctikzset{bipoles/thickness=1}

\newlength{\dotsclength}
\setlength{\dotsclength}{\widthof{\footnotesize$\dotsc$}}

\usetikzlibrary{decorations.markings,chains}
\tikzset{threedots/.style={decorate,decoration={markings,mark=between			positions 0.25 and 0.75 step 0.25 with {\node[circle,draw=black,fill=black,inner sep=0pt,minimum	size=1pt,thin]{};}}}}

\usepackage{pgfplots}
\pgfplotsset{compat=1.14}
\pgfplotsset{every axis/.append style={line width=0.75pt,tick style={line width=0.25pt},grid style={line width=0.5pt}}}

\usetikzlibrary{arrows.meta,calc,positioning}

\usepackage[outline]{contour}
\contourlength{0.8pt}

\tikzset{double_arrow/.style={-{Triangle[length=3.9pt,white,width=5pt]},double distance=5pt,thick,postaction={decorate,decoration={markings,mark=at position 1 with {\arrow[scale=2.5,thin]{Straight Barb}}}}
}}

\tikzset{vh path/.style={to path={|- (\tikztotarget)},DSP lines},hv path/.style={to path={-| (\tikztotarget)},DSP lines}}

\tikzset{adder/.style={circle,minimum size=.25cm,inner sep=0pt,draw=black,very thick,execute at end node={$\textbf{+}$}}}

\tikzset{filter/.style={rectangle,inner sep=2pt,minimum height=0.6cm,draw=black,very thick}}

\usetikzlibrary{backgrounds}
\tikzstyle{thicker}=[line width=1pt]

\usetikzlibrary{external}
\tikzexternalize[prefix=external/]

\definecolor{capcolor}{RGB}{237,99,105}%
\definecolor{rhypcolor}{RGB}{0,77,128}%
\definecolor{rrecipcolor}{HTML}{edc769}%
\definecolor{rassumedcolor}{RGB}{79,185,255}%
\definecolor{rbothcolor}{rgb}{0.46600,0.67400,0.18800}%

\definecolor{tempcolor}{rgb}{0.00000,0.44700,0.74100}%
\colorlet{histogramcolor}{tempcolor!60!white}

\tikzset{capmark/.style={mark=triangle}}
\tikzset{rhypmark/.style={mark=x}}
\tikzset{rrecipmark/.style={mark=diamond}}
\tikzset{rassumedmark/.style={mark=o}}
\tikzset{rbothmark/.style={mark=x}}

\usetikzlibrary{intersections}
}

\DeclareSIUnit\decibelwatt{dBW}

\newcommand{\varHDL}{\matt H}
\newcommand{\varsigmathetaDL}{\sigma_{\vartheta}}
\newcommand{\varRetaDL}{\matt R_{\eta}}
\newcommand{\varDDL}{\matt D}
\newcommand{\varBDL}{\matt B}
\newcommand{\varQDL}{\matt Q}
\newcommand{\varsigmaetaDL}{\sigma_{\eta}}

\newcommand{\varhDL}{\vect h}

\newcommand{\varBhatDL}{\hat{\matt B}}
\newcommand{\varhhatDL}{\hat{\vect h}}
\newcommand{\varHhatDL}{\hat{\vect H}}

\newcommand\varRetahatDL{\hat{\matt R}_{\eta}}
\newcommand\varthetahatDL{\hat{\vect \vartheta}}
\newcommand\varRthetahatDL{\hat{\matt R}_{\vartheta}}

\begin{document}
\title{On Reciprocity in Physically Consistent TDD~Systems with Coupled Antennas}%
\author{Tobias~Laas,~\IEEEmembership{Graduate Student Member,~IEEE,}
	    Josef~A.~Nossek,~\IEEEmembership{Life~Fellow,~IEEE,} 
	    Samer~Bazzi,
	    and~Wen~Xu,~\IEEEmembership{Senior~Member,~IEEE}%
	    \thanks{This work has been partly performed in the framework of the Horizon 2020 project ONE5G (ICT-760809) receiving funds from the European Union. This paper was presented in part at the 21st International ITG Workshop on Smart Antennas (WSA), Berlin, Germany, March 2017.}%
	\thanks{T.~Laas is with the German Research Center, Huawei Technologies Duesseldorf GmbH, 80992 Munich, Germany, and also with the Department of Electrical and Computer Engineering, Technical University of Munich, 80333 Munich, Germany (e-mail: tobias.laas@tum.de).}%
    \thanks{J.~A.~Nossek is with the Department of Electrical and Computer Engineering, Technical University of Munich, 80333 Munich, Germany, and also with the Department of Teleinformatics Engineering, Federal University of Ceará, Fortaleza, Brazil (e-mail: josef.a.nossek@tum.de).}%
	\thanks{S. Bazzi and W. Xu are with the German Research Center, Huawei Technologies Duesseldorf GmbH, 80992 Munich, Germany (e-mail: samer.bazzi@huawei.com; wen.xu@ieee.org).}%
}%
\maketitle%
\begin{tikzpicture}[remember picture,overlay]
\node[yshift=2cm] at (current page.south){\parbox{\textwidth}{\footnotesize \textcopyright\ 2020 IEEE. Personal use of this material is permitted. Permission from IEEE must be obtained for all other uses, in any current or future media, including reprinting/republishing this material for advertising or promotional purposes, creating new collective works, for resale or redistribution to servers or lists, or reuse of any copyrighted component of this work in other works.}};
\node[yshift=-1cm] at (current page.north){\parbox{\textwidth-2cm}{\footnotesize This is the accepted version of the following article: 
T.~Laas, J.~A. Nossek, S.~Bazzi, and W.~Xu, ``On reciprocity in physically consistent TDD systems with coupled antennas,'' \emph{IEEE Trans. Wireless Commun.}, 2020, \doi{10.1109/TWC.2020.3003414}.}};
\end{tikzpicture}%
\begin{abstract}
	We consider the reciprocity of the information-theoretic channel of Time Division Duplex (TDD) Multi-User-Multiple Input Multiple Output (MU-MIMO) systems in the up- and downlink. Specifically, we assume that the transmit and receive chains are reciprocal. We take the mutual coupling between the antenna elements at the base station and at the mobiles into account. Mutual coupling influences how to calculate transmit power and noise covariance. The analysis is based on the Multiport Communication Theory, which ensures that the information-theoretic model is consistent with physics. It also includes a detailed noise model. We show that due to the coupling, the information-theoretic up- and downlink channels do not fulfill the ordinary reciprocity relation, even if the input-output relation of the transmit voltage sources and the receive load voltages, i.e., the channel which is estimated with the help of pilot signals in the uplink, is reciprocal. This is a fundamental effect that is not considered otherwise. We show via Monte Carlo simulations that both, using the ordinary reciprocity relation, and not taking the coupling into account, significantly decreases the ergodic rates in single-user and the ergodic sum rates in multi-user systems. 
\end{abstract}%
\section{Introduction}
\IEEEPARstart{C}{urrently} deployed wireless standards such as LTE only employ a small number of antennas at the mobiles and at the base station. It is expected that to accommodate further growth of the amount of transferred data, a significantly larger number of antennas needs to be employed at the base station.
In order to exploit the degrees of freedom provided by the antennas, the base station requires channel state information (CSI). The amount of CSI increases with the number of antennas. In frequency division duplex (FDD) mode, the base station can usually acquire downlink CSI by sending pilot signals, letting the mobiles estimate the CSI and feed back the estimate. 
The advantage of time division duplex (TDD) mode is that the base station can reuse CSI from the uplink, as the physical channel is reciprocal~\cite{LarssonMMimoCommag}. The uplink CSI can be acquired with less pilot overhead than the downlink CSI if there are in total fewer antennas at the mobiles than at the base station.

In practical systems, the transmit (Tx) and receive (Rx) RF chains are usually not identical, i.e., up- and downlink channels are not reciprocal. Reciprocity calibration is used to take this into account~\cite{PetermannJournal,KaltenbergerMobileSummit2010,SamerWenOTACali2016,VieiraGlobecom,VieiraJournal,Wei}.
In some of these papers, the mutual coupling between the antenna elements of the same array is leveraged for the calibration process. But they do not take into account that mutual coupling itself has an impact on the reciprocity relation of the up- and downlink channel matrices in the information-theoretic model. Here we assume that one of the methods for calibrating the RF chains is applied such that those in the uplink and those in the downlink are made equal in the DSP part of the system.

We will show that there is another fundamental source changing the reciprocity relation, namely mutual coupling, using the so-called Multiport Communication Theory~\cite{IvrlacNossekTowardaTheory,IvrlacNossekMultiportCommTheory}. This theory is in turn based on a circuit-theoretical description using impedance matrices and provides a way to model the system consistently with physics. The circuit model can equivalently be described using scattering matrices, similar to the models~\cite{WallaceJensen,WaldschmidtTVT}, which to the authors' best knowledge were the first to take into account that mutual coupling changes the transmit power. The model in~\cite{WallaceJensen} was later extended by a detailed amplifier noise model~\cite{MorrisJensenNetworkModel}.
Two similar noise models that contain both antenna and amplifier noise are introduced in \cite{IvrlacNossekTowardaTheory,IvrlacNossekMultiportCommTheory}, and~\cite{DomiziloiNoise}. The key contributions of~\cite{IvrlacNossekTowardaTheory} and \cite{IvrlacNossekMultiportCommTheory} are that it merges the circuit model and the noise model, and that it introduces a systematic mapping from the circuit-based models to the usual information-theoretic model. All of these models only consider a one-way link, so it is applied to both up- and downlink in analyzing reciprocity. In conventionally modeled systems, the information-theoretic ordinary (pseudo-physical) reciprocity relation $\varHDL=\matt H_\mathrm{UL}^T$ is employed, but due to mutual coupling, this relation does not hold, but rather a new physically consistent reciprocity relation. This is because mutual coupling influences how to calculate transmit power and noise covariance.

\begin{figure*}[!t]
	\centering
	\ifbool{pdffigures}{%
		\includegraphics{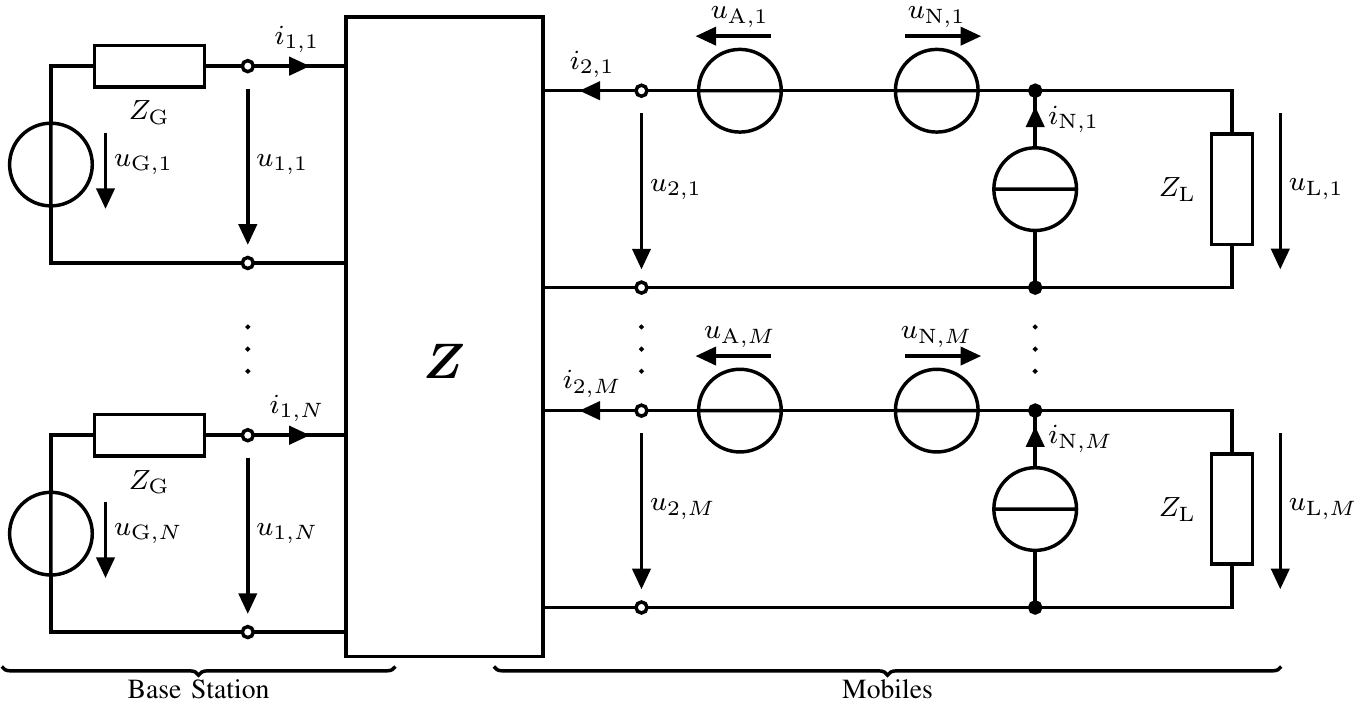}%
	}{%
		\tikzsetnextfilename{impedancenetworkDL}%
		\input{impedancenetworkDL}%
	}%
	\caption{Circuit model in the downlink.}%
	\label{fig:circuitmodelDL}%
\end{figure*}%
The rest of the paper is organized as follows: first we review a simple circuit-theoretic model in Section~\ref{sec:multiportcommtheory}, then we consider the reciprocity of the information-theoretic channel and show how to take it into account in Section~\ref{sec:unusualrecip}. We analyze the effect on the radiated power and on the (sum) rates in the single user multiple input single output (SU-MISO), single user multiple input multiple output (SU-MIMO), multi user MISO (MU-MISO) and MU-MIMO downlink, first theoretically, see Section~\ref{sec:capandrates}, and second in simulation in independent and identically distributed (i.i.d.) channels and in QuaDRiGa~\cite{QuaDRiGA,QuaDRiGATAP} channels, see Sections~\ref{sect:sim} and~\ref{sec:simquadriga}. The SU-MISO case was presented in part in a conference paper~\cite{WSAPaper}. Conclusions follow in Section~\ref{sec:conclusions}.

\textit{Notation}: lowercase bold letters denote vectors, uppercase bold letters matrices. $a_m$ denotes the $m$th element of $\vect a$. $\matt A^T, \matt A^\ast, \matt A^H, \abs*{\matt A}, \norm{\matt A}_\mathrm{F}, \tr(\matt A)$ and $\diag(\matt A)$ correspond to the transpose, the complex conjugate, the Hermitian, the determinant, the Frobenius norm, the trace and the matrix whose diagonal elements are equal to those of $\matt A$ and whose other entries are zero. $\vect 0$ and $\matt I$ denote zero vector and identity matrix. $\mathcal{N}_\mathbb{C}(\vect \mu, \matt R)$ denotes a proper complex Gaussian distribution with mean $\vect \mu$ and covariance $\matt R$.  $\E[ X ]$ denotes the expectation of the random variable $X$.

\section{Multiport Communication Theory}
\label{sec:multiportcommtheory}%
\subsection{Circuit-Theoretic Model}
We focus on a simple circuit model (Fig.~\ref{fig:circuitmodelDL}) for the fading channel by assuming the that the fading is flat within a narrowband (group of) subcarrier(s) of a multicarrier system\footnote{This is the standard assumption in most of the MIMO literature.}, similar to the ones in~\cite{IvrlacNossekTowardaTheory},~\cite[Fig.~9]{IvrlacNossekMultiportCommTheory} and~\cite{SCC2010Ivrlac,IvrlacWSA2016}, where simple means that as in~\cite{SCC2010Ivrlac,IvrlacWSA2016}, we omit the lossless decoupling and impedance matching network (DMNs), because in massive MIMO systems, they could be almost impossible to implement. But as in~\cite{IvrlacNossekTowardaTheory,IvrlacNossekMultiportCommTheory} and \cite{IvrlacWSA2016}, we also consider the thermal noise of the antennas. 

The signal generation at the transmitter is modeled as a linear voltage source $u_{\mathrm{G},n}$ with internal impedance $Z_\mathrm{G}$ per antenna. The antennas are assumed to be lossless~\cite{IvrlacNossekMultiportCommTheory} and their coupling and the physical channel are modeled jointly by an impedance matrix $\matt Z$. 
At the receivers, each hardware chain is modeled by an impedance $Z_\mathrm{L}$ and several noise sources, which we will come back to later. 

Let there be in total $N$ antennas at the transmitter(s) and $M$~at the receiver(s). As antennas and the physical channel are reciprocal~\cite{SchelkunoffFriis}, the system described by the impedance matrix $\matt Z~\in~\mathbb{C}^{(N+M)\times(N+M)} \cdot \si{\ohm}$ is reciprocal as well, i.e.,
\begin{equation}
\label{equ:zrecip}
\matt Z = \matt Z^T.
\end{equation}
It is partitioned into four blocks~\cite{IvrlacNossekTowardaTheory}: the transmit and receive impedance matrices $\matt Z_{11} \in \mathbb{C}^{N\times N}\cdot\si{\ohm}$ and $\matt Z_{22} \in \mathbb{C}^{M\times M}\cdot \si{\ohm}$, and the mutual impedance matrices $\matt Z_{21} \in \mathbb{C}^{M\times N}\cdot \si{\ohm}$ and $\matt Z_{12} \in \mathbb{C}^{N\times M}\cdot \si{\ohm}$ such that 
\begin{equation}
\begin{bmatrix}
\vect u_1\\
\vect u_2
\end{bmatrix}
= \underbrace{\begin{bmatrix} \matt Z_{11} & \matt Z_{12}\\
\matt Z_{21} & \matt Z_{22}
\end{bmatrix}}_{\matt Z} \begin{bmatrix} \vect i_1 \\ \vect i_2
\end{bmatrix},
\end{equation}
where $\vect u_1 \in \mathbb{C}^{N} \cdot \si{\volt}, \vect i_1 \in \mathbb{C}^{N} \cdot \si{\ampere}, \vect u_2\in \mathbb{C}^{M} \cdot \si{\volt}, \vect i_2\in \mathbb{C}^{M} \cdot \si{\ampere}$ are the port voltages and currents at the transmitter and receiver side~\cite{IvrlacNossekTowardaTheory} (see~Fig.~\ref{fig:circuitmodelDL}). All voltages and currents in this paper are rms values of complex envelopes.

Let us consider the relation between the generator and load voltages $\vect u_\mathrm{G}~\in~\mathbb{C}^{N} \cdot \si{\volt}$ and $\vect u_\mathrm{L}~\in~\mathbb{C}^{M} \cdot \si{\volt}$. Compared to~\cite{IvrlacNossekTowardaTheory}, the relation between voltages and currents at the generator side simplifies to
\begin{equation}
\label{equ:ugrelbasic}
\vect u_\mathrm{G} = \vect u_1 + Z_\mathrm{G} \vect i_1.
\end{equation}
Using the unilateral approximation $\norm{\matt Z_{12}}_\mathrm{F}~\ll~\norm{\matt Z_{11}}_\mathrm{F}$ \cite{IvrlacNossekTowardaTheory}, whereby we assume that the attenuation of the channel is so high that the currents in the antennas at the receivers do not influence the transmitter, we have~\cite{IvrlacNossekTowardaTheory}
\begin{equation}
\label{equ:u1i1unilateral}
\vect u_1 = \matt Z_{11} \vect i_1.
\end{equation}
According to the superposition theorem,
\begin{equation}
\label{equ:voltagerelation}
\begin{split}
\vect u_\mathrm{L} &= \underbrace{\vect u_\mathrm{L}\rvert_\mathrm{nf}}_{\textrm{noise-free}}+ \underbrace{\vect u_\mathrm{L}\rvert_\mathrm{sf}}_{\textrm{signal-free}}\breakiftwocolumn 
\aligniftwocolumn=\vect u_\mathrm{L}\rvert_\mathrm{nf} + \sqrt{R_\mathrm{L}}\vect \eta, \quad R_\mathrm{L} \coloneqq \Real(Z_\mathrm{L}),
 \end{split}
\end{equation}
where $\vect \eta$ describes the noise and will be given in \eqref{equ:defeta}.
\begin{figure*}[!t]
	\centering
	\ifbool{pdffigures}{%
		\includegraphics{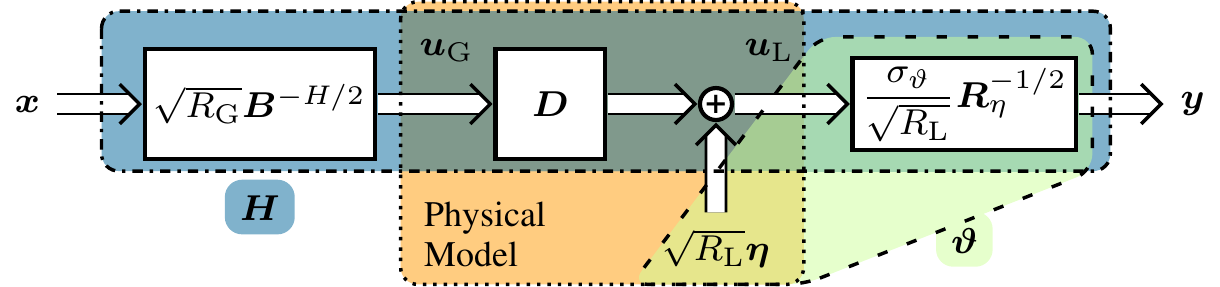}%
	}{%
		\tikzsetnextfilename{physinfoblkdiagram}%
		\input{physinfoblkdiagram}%
	}%
	\caption{System model showing the relation between the physical and the information-theoretic model.}%
	\label{fig:physinfoblkdiagram}%
\end{figure*}%

The excitation in the noise-free case is caused by $\vect u_\mathrm{G}$, and in the signal-free case by the noise sources.
We use the same noise model as in~\cite{IvrlacNossekMultiportCommTheory}, which distinguishes between the extrinsic noise $\vect u_\mathrm{A}\in \mathbb{C}^{M} \cdot \si{\volt}$ produced by the antennas in thermal equilibrium, and the intrinsic noise, which stems mainly from the LNAs (but also from other components)~\cite{IvrlacNossekTowardaTheory}, which can be jointly modeled as noisy two-ports. There is an equivalent model~\cite{NoisyFourpoles} for each of the noisy two-ports consisting of a noiseless two-port with a voltage and a current noise source, $u_{\mathrm{N},m}, i_{\mathrm{N},m}$, at its input, which model the intrinsic noise. The SNR at the input and the output of the noiseless two-port is the same and thus it is sufficient to only consider the input port in the model~\cite{IvrlacNossekMultiportCommTheory}.

The noise distributions are modeled as~\cite{IvrlacNossekTowardaTheory}
\begin{align}
\label{equ:noisedistribution}
\vect u_\mathrm{A} &\sim \mathcal{N}_\mathbb{C}(\vect 0\,\si{\volt}, \matt R_\mathrm{A}), \quad  \matt R_\mathrm{A} = 4 k_\mathrm{B}  T_\mathrm{A} \Delta f \Real(\matt Z_{22}),\\
\nonumber \vect u_\mathrm{N} &\sim \mathcal{N}_\mathbb{C}(\vect 0\,\si{\volt}, \sigma_u^2 \matt I), \qquad 
\vect i_\mathrm{N} \sim \mathcal{N}_\mathbb{C}(\vect 0\,\si{\ampere}, \sigma_i^2 \matt I)
\end{align}
for some $\sigma_u > \SI{0}{\volt}, \sigma_i >\SI{0}{\ampere}$, where $k_\mathrm{B}$ is the Boltzmann constant, $\Delta f$ is the noise bandwidth and $T_\mathrm{A}$ is the noise temperature of the antennas.
In the noise-free case,
\begin{align}
\label{equ:ulequinter}
\vect u_\mathrm{L}\rvert_\mathrm{nf} &= \vect u_2\rvert_\mathrm{nf} = \matt Z_{21} \vect i_1 + \matt Z_{22} \vect i_2\rvert_\mathrm{nf} = - Z_\mathrm{L} \vect i_2\rvert_\mathrm{nf}.
\shortintertext{Combining \eqref{equ:ugrelbasic}, \eqref{equ:u1i1unilateral} and \eqref{equ:ulequinter} leads to }
\label{equ:defd}
\begin{split}
\vect u_\mathrm{L}\rvert_\mathrm{nf} &= \matt D \vect u_\mathrm{G},\breakiftwocolumnquad
\matt D \aligniftwocolumn= Z_\mathrm{L} (\matt Z_{22} + Z_\mathrm{L} \matt I )^{-1} \matt Z_{21} ( \matt Z_{11}+ Z_\mathrm{G} \matt I )^{-1}.
\end{split}
\end{align}
In the signal-free case, the intrinsic noise sources $\vect u_\mathrm{N}$ and $\vect i_\mathrm{N}$ are assumed to be uncorrelated with the extrinsic noise $\vect u_\mathrm{A}$. $u_{\mathrm{N},m}$ and $i_{\mathrm{N},m}$ are correlated with the correlation coefficient~\cite{IvrlacNossekTowardaTheory}
\begin{equation}
\label{equ:noiserho}
\rho = \frac{\E [u_{\mathrm{N},m} i_{\mathrm{N},m}^\ast]}{\sigma_u\sigma_i} \quad \forall m.
\end{equation}
Consider the following two equations that follow from Kirchhoff's voltage and current law:
\begin{gather}
-\vect u_\mathrm{A}+\vect u_\mathrm{N}+\vect u_\mathrm{L}\rvert_\mathrm{sf} = \matt Z_{22} \vect i_2\rvert_\mathrm{sf},\\
\vect i_2\rvert_\mathrm{sf} = \vect i_\mathrm{N} -Z_\mathrm{L}^{-1} \vect u_\mathrm{L}\rvert_\mathrm{sf}.
\end{gather}
Eliminating $\vect i_2\rvert_\mathrm{sf}$ and solving for $\vect u_\mathrm{L}\rvert_\mathrm{sf}$ gives the relation between $\vect \eta$ and the noise sources
\begin{equation}
\label{equ:defeta}
\vect \eta = \frac{\vect u_\mathrm{L}\rvert_\mathrm{sf}}{\sqrt{R_\mathrm{L}}} = \frac{Z_\mathrm{L}}{\sqrt{R_\mathrm{L}}} ( \matt Z_{22} + Z_\mathrm{L} \matt I )^{-1} (\vect u_\mathrm{A}-\vect u_\mathrm{N}+\matt Z_{22} \vect i_\mathrm{N}).
\end{equation}
Together with \eqref{equ:noisedistribution} and \eqref{equ:noiserho}, the noise covariance matrix can be computed as
\begin{equation}
\begin{split}
\matt R_\eta &= \E [\vect \eta \vect \eta^H ] = \frac{\abs{Z_\mathrm{L}}^2}{R_\mathrm{L}} ( \matt Z_{22} + Z_\mathrm{L} \matt I )^{-1} \matt Q ( \matt Z_{22} + Z_\mathrm{L} \matt I )^{-H},\\
\matt Q&= \sigma_u^2 \matt I + \sigma_i^2 \matt Z_{22} \matt Z_{22}^\ast  - 2 \sigma_u \sigma_i \Real(\rho^\ast \matt Z_{22})  + \matt R_\mathrm{A}.
\end{split}\raisetag{13pt}
\end{equation}
The transmit power in the physical model can be computed as
\begin{equation}
\label{equ:defb}
\begin{split}
P_\mathrm{T} &= \E[\Real (\vect i_1^H \vect u_1)] = \frac{\E[\vect u_\mathrm{G}^H \matt B \vect u_\mathrm{G}]}{R_\mathrm{G}}, \quad R_\mathrm{G} \coloneqq \Real(Z_\mathrm{G}),\\
\matt B &= R_\mathrm{G} \parens*{\matt Z_{11} + Z_\mathrm{G} \matt I}^{-H} \Real(\matt Z_{11}) \parens*{\matt Z_{11} + Z_\mathrm{G} \matt I }^{-1},
\end{split}\raisetag{13pt}
\end{equation}
where we have used \eqref{equ:ugrelbasic} and where $\matt B$ is the so-called power-coupling matrix~\cite{IvrlacNossekTowardaTheory}. Then the complete physical model is
\begin{equation}
\label{equ:physicalmodel}
\begin{split}
\vect u_\mathrm{L} = \matt D \vect u_\mathrm{G} + \sqrt{R_\mathrm{L}} \vect \eta, \quad \vect \eta \sim \mathcal{N}_\mathbb{C}( \vect 0\,\si{\sqrt{\watt}}, \matt R_\eta),\breakiftwocolumnquad \quad P_\mathrm{T} = \frac{\E[\vect u_\mathrm{G}^H \matt B \vect u_\mathrm{G}]}{R_\mathrm{G}}.
\end{split}
\end{equation}

\subsection{Information-Theoretic Model}
Consider the typical information-theoretic model (e.g., ~\cite[Ch.~1]{MIMOWComm})
\begin{equation}
\label{equ:itmodel}
\begin{split}
\vect y &= \matt H \vect x + \vect \vartheta, \quad \vect \vartheta \sim \mathcal{N}_\mathbb{C}( \vect 0\,\si{\sqrt{\watt}}, \sigma_\vartheta^2 \matt I), \quad \sigma_\vartheta>\SI{0}{\sqrt{\watt}},\breakiftwocolumnquad P_\mathrm{T} \aligniftwocolumn= \E [\vect x^H \vect x ],
\end{split}\raiseintwocolumn{12pt}
\end{equation}
which allows existing techniques and results for capacity and achievable rates to be easily draw on. In order to get a physically consistent information-theoretic model, we need to ensure that the transmit power $P_\mathrm{T}$ and the noise covariance are consistent with the physical model \eqref{equ:physicalmodel}. This can be achieved by a linear mapping from $\vect u_\mathrm{G}$ and $\vect u_\mathrm{L}$ to $\vect x$ and $\vect y$,
\begin{align}
\label{equ:consistentxug}
\vect x &= \frac{1}{\sqrt{R_\mathrm{G}}}\matt B^{H/2} \vect u_\mathrm{G}, \quad &\mathrm{s.t.}& \quad \matt B = \matt B^{1/2} \matt B^{H/2},\\
\label{equ:consistentyul}
\vect y &= \frac{\sigma_\vartheta}{\sqrt{R_\mathrm{L}}}\matt R_\eta^{-1/2} \vect u_\mathrm{L},
\end{align}
as shown in~\cite{IvrlacNossekTowardaTheory,IvrlacNossekMultiportCommTheory} and leads to the system model shown in  Fig.~\ref{fig:physinfoblkdiagram}. Throughout the paper, we assume that matrix square roots in general fulfill a condition similar to \eqref{equ:consistentxug}. The expressions are not exactly the same as in~\cite{IvrlacNossekTowardaTheory,IvrlacNossekMultiportCommTheory}, because $\matt B^{1/2}$ is not unique and in these two papers, only $\matt B^{1/2}$ that are Hermitian are considered. We choose
\begin{align}
\label{equ:defbsqrt}
\begin{split}
\matt B^{1/2} &= \sqrt{R_\mathrm{G}}\parens*{\matt Z_{11} + Z_\mathrm{G} \matt I}^{-H} \Real(\matt Z_{11})^{1/2}\breakiftwocolumnquad\aligniftwocolumn\mathrm{s.t}\quad \Real(\matt Z_{11})=\Real(\matt Z_{11})^{1/2}\Real(\matt Z_{11})^{1/2},
\end{split}\\
\matt R_\eta^{1/2} &= \frac{Z_\mathrm{L}}{\sqrt{R_\mathrm{L}}} \parens*{\matt Z_{22} + Z_\mathrm{L} \matt I}^{-1} \matt{Q}^{1/2}.
\end{align}
This leads to the information-theoretic channel
\begin{equation}
\label{equ:defh}
\begin{split}
\matt H &= \sigma_\vartheta \frac{\sqrt{R_\mathrm{G}}}{\sqrt{R_\mathrm{L}}} \matt R_\eta^{-1/2} \matt D \matt B^{-H/2} \breakiftwocolumn \aligniftwocolumn
        = \sigma_\vartheta \matt Q^{-1/2} \matt Z_{21} \Real(\matt Z_{11})^{-1/2},
\end{split}
\end{equation}
which captures the physical context~\cite{IvrlacNossekTowardaTheory,IvrlacNossekMultiportCommTheory}.
$\sigma_\vartheta$ is an arbitrary scaling, but to ensure that the sum noise powers in the physical and information-theoretic models are the same, i.e.,
\begin{equation}
\E[\vect \vartheta^H \vect \vartheta] = \E[\vect \eta^H \vect \eta]
\end{equation}
holds, let
\begin{equation}
\label{equ:noisenorm}
\sigma_\vartheta^2=\frac{\tr(\matt R_\eta )}{M}.
\end{equation}

\subsection{Neglecting the Mutual Coupling}
\label{subsec:neglmuimp}
There are three matrices that characterize the information-theoretic channel $\matt H$, namely $\matt B$, $\matt D$ and $\matt R_\eta$. The matrix $\matt D$ can be estimated with the help of pilot symbols. Independent of whether the mutual coupling is neglected or not, the estimate related to perfect CSI knowledge is always the same $\matt D$. It is the only matrix of the three that is time-variant due to user mobility. The other two are time-invariant. $\matt B$ is a function of $Z_\mathrm{G}$ and $\matt Z_{11}$ -- or equivalently the scattering parameters
-- which can be determined by off-line modeling, simulation or measurement of the antenna arrays, including the front/back end of the RF chains.  In many publications, mutual coupling is ignored, meaning that  $\matt B$ and $\matt R_\eta$ are assumed to be diagonal or scaled identity matrices (see Section~\ref{sec:capandrates}). Acquiring $\matt R_\eta$ is further discussed in the following section.

\section{Reciprocity of the Information-Theoretic Channel}
\label{sec:unusualrecip}
From now on, we relate the models presented in the previous section to the downlink (Fig.~\ref{fig:circuitmodelDL}). The uplink uses a similar model, but with the impedance matrix $\matt Z^T$ and with the noise sources at the base station. $\matt Z_{11}$ describes the antennas at the base station and $\matt Z_{22}$ those at the mobiles. In the following sections, we will assume that $Z_\mathrm{G}=Z_\mathrm{L}$, as $Z_\mathrm{G}\ne Z_\mathrm{L}$ will be compensated due to reciprocity calibration. Then, due to the symmetry between ``1'' and ``2'' in \eqref{equ:defd} and as $\matt Z_{21} = \matt Z_{12}^T$ (see \eqref{equ:zrecip}), the noiseless relation between generator and load voltage, $\varDDL$, is reciprocal, i.e.,
\begin{equation}
\label{equ:drecip}
\matt D_\mathrm{UL}^T = \varDDL = Z_\mathrm{L} \parens*{\matt Z_{22} + Z_\mathrm{L} \matt I}^{-1} \matt Z_{21} \parens*{\matt Z_{11}+ Z_\mathrm{G} \matt I}^{-1}.
\end{equation}
However, there is no such symmetry in \eqref{equ:defh}, but
\begin{align}
\label{equ:hdl}
\begin{split}\varHDL &=\varsigmathetaDL \varRetaDL^{-1/2} \varDDL \varBDL^{-H/2} \breakiftwocolumn \aligniftwocolumn = \varsigmathetaDL \varQDL^{-1/2} \matt Z_{21} \Real(\matt Z_{11})^{-1/2}
\end{split}
\shortintertext{and}
\label{equ:hul}
\begin{split}\matt H_\mathrm{UL} &= \sigma_{\vartheta,\mathrm{UL}}  \matt R_{\eta,\mathrm{UL}}^{-1/2} \matt D_\mathrm{UL} \matt B_\mathrm{UL}^{-H/2}\breakiftwocolumn \aligniftwocolumn =\sigma_{\vartheta,\mathrm{UL}} \matt Q_\mathrm{UL}^{-1/2} \matt Z_{12} \Real(\matt Z_{22})^{-1/2}
\end{split}
\end{align}
hold, so the information-theoretic downlink and uplink channels are not reciprocal in the ordinary way, i.e., $\matt H_\mathrm{UL}^T \ne \varHDL$. Although $\varDDL$ is reciprocal, in general, a different reciprocity relation is introduced by whitening the noise coupling between the antennas and by maintaining the physical consistency of the transmit power, see \eqref{equ:consistentxug} and \eqref{equ:consistentyul}. This physically consistent reciprocity relation 
\begin{equation}
\label{equ:hdlhulrelation}
\varHDL = \frac{\varsigmathetaDL}{\sigma_{\vartheta,\mathrm{UL}}}\varRetaDL^{-1/2} \matt B_{\mathrm{UL}}^{\ast/2} \matt H_\mathrm{UL}^T \matt R_{\eta,\mathrm{UL}}^{T/2} \varBDL^{-H/2}
\end{equation}
is obtained by comparing \eqref{equ:hdl} and \eqref{equ:hul}. If the base station wants to reuse the CSI estimated in the uplink for the downlink, it needs to use this physically consistent reciprocity relation for the information-theoretic channel.

Consider that the base station acquires CSI in the uplink by estimating $\matt D_\mathrm{UL}$ -- instead of $\matt H_\mathrm{UL}$ -- from the mobile(s)' pilot symbols. To make the downlink physically consistent at the base station, i.e., to apply \eqref{equ:consistentxug}, it needs to know $\varBDL^{-H/2}$, anyway, so that it can compute $\matt D_\mathrm{UL}^T \varBDL^{-H/2}$ without any further information, although compared to \eqref{equ:hdl}, there remains the unknown factor $\varsigmathetaDL \varRetaDL^{-1/2}$.

Let us simplify the model for the MU-MISO downlink and uplink, i.e., when each mobile has a single antenna. We assume that the distance between different mobiles is large with respect to the wavelength. For large distances, the coupling reduces inversely with the  distance~\cite{IvrlacNossekMultiportCommTheory}, so it goes to zero and $\matt Z_{22}$ becomes diagonal. Furthermore, we assume identical antenna impedances $Z_\mathrm{A}$ at the mobiles, i.e.,
\begin{equation}
\label{equ:diagonalzaz22}
\matt Z_{22} = Z_\mathrm{A} \matt I.
\end{equation}
Then the downlink information-theoretic channel simplifies to
\begin{equation}
\label{equ:misoqsimplified}
\begin{split}
\varRetaDL \aligniftwocolumn= \varsigmaetaDL^2 \matt I, \quad \varsigmaetaDL^2 = \frac{\abs{Z_\mathrm{L}}^2}{R_\mathrm{L}\abs{Z_\mathrm{A} + Z_\mathrm{L}}^2} \sigma_q^2, \quad \matt \varQDL = \sigma_q^2 \matt I,\breakiftwocolumnquad
\varHDL \aligniftwocolumn= \matt D_\mathrm{UL}^T  \varBDL^{-H/2},
\end{split}
\end{equation}
i.e., due to \eqref{equ:noisenorm}, there is no unknown factor $\varsigmathetaDL \varRetaDL^{-1/2}$ in this scenario. The uplink information-theoretic channel simplifies to
\begin{equation}
\label{equ:misoulsimplified}
\matt H_\mathrm{UL} = \sigma_{\vartheta,\mathrm{UL}}\matt R_{\eta,\mathrm{UL}}^{-1/2} \matt D_\mathrm{UL}  \frac{Z_\mathrm{A}+Z_\mathrm{G}}{\sqrt{R_\mathrm{G} \Real(Z_\mathrm{A})}}.
\end{equation}
Also in this case, $\varHDL$ and $\matt H_\mathrm{UL}$ are not reciprocal in the ordinary way. 

Note that if the mobiles have more than one antenna, i.e., in MU-MIMO systems, $\matt Z_\mathrm{22}, \varQDL, \varRetaDL$ and $\matt B_\mathrm{UL}$ are block diagonal, since we assume that there is no coupling between different mobiles. The noise covariance of multi-antenna mobiles is also a matrix instead of a scalar. Therefore, even if \eqref{equ:noisenorm} is taken into account, $\matt D_\mathrm{UL}^T  \varBDL^{-H/2} \ne \varHDL$ in general.
One solution in practice might be to create a database of noise covariance matrices corresponding to different models of mobiles for the base station. As in conventionally modeled systems, the base station needs feedback from the mobiles about their $\mathrm{SNR}$ and needs to deal with $\matt R_\eta$, which is not known perfectly.

\section{Capacities and Rates not Taking the Physical Reciprocity or the Mutual Coupling into Account}
\label{sec:capandrates}
\begin{table*}[!t]
	\renewcommand{\arraystretch}{1.3}
	\caption{Overview of the Transmit Strategies}%
	\label{tab:schemes}%
	\centering%
	\begin{tabular}{p{0.145\linewidth}|p{0.25\linewidth}|p{0.25\linewidth}|p{0.25\linewidth}|}
		&``cap'' & ``recip'' & ``hyp''\\
		\hline 
		Rates &
		$C, R_\mathrm{lin}$ &
		$R_\mathrm{recip}, R_\mathrm{recip,lin}$ &
		$R_\mathrm{hyp}, R_\mathrm{hyp,lin}$\\
		\hline 
		Motivation & 
		Capacity achieving strategy.%
		& Naive use of the Multiport Communication Theory.
		&
		Conventionally modeled systems.\\
		\hline
		Description &
		The base station uses the information-theoretic model \eqref{equ:itmodel}, as it leads to an easy to use channel model~\cite{IvrlacNossekMultiportCommTheory}, \bfseries and uses the physically consistent reciprocity relation \eqref{equ:hdlhulrelation}. &
		The base station uses the information-theoretic model \eqref{equ:itmodel}, as it leads to an easy to use channel model~\cite{IvrlacNossekMultiportCommTheory}, \bfseries but assumes that the ordinary reciprocity relation holds. &
		The base station uses the information-theoretic model ignoring the mutual coupling (see \eqref{equ:transformingcoupling}, \eqref{equ:defPTp}, \eqref{equ:mimoitchannnelmismatch} and \eqref{equ:yhatmapping}) and the ordinary reciprocity relation.\\
		\hline
		Procedure &
		\begin{algorithmic}[1]
			\State Estimate $\matt D_\mathrm{UL}$.
			\State Compute $\varHDL$ using \eqref{equ:hul} and \eqref{equ:hdlhulrelation} (or \eqref{equ:misoqsimplified}).
			\State Apply the optimal transmit strategy based on $\varHDL$ and $P=P_\mathrm{T}$.
		\end{algorithmic}
		&
		\begin{algorithmic}[1]
			\State Estimate $\matt D_\mathrm{UL}$.
			\State Compute $\matt H_\mathrm{UL}$ using \eqref{equ:hul}.
			\State Apply the optimal transmit strategy based on $\matt H_\mathrm{UL}^T$ (corresponding to $\vect h_\mathrm{UL}^\ast$ for SU-MISO) and $P=P_\mathrm{T}$.
		\end{algorithmic}
		&			
		\begin{algorithmic}[1]
			\State Estimate $\matt D_\mathrm{UL}$.
			\State Compute $\varHhatDL$ using \eqref{equ:drecip} and \eqref{equ:mimoitchannnelmismatch}.
			\State Apply the optimal transmit strategy based on $\varHhatDL^\prime$ (see \eqref{equ:hhatDLdef}) and $P=P_\mathrm{T,p}$.
		\end{algorithmic}
		\\
		\hline Channel that is transmitted over in the downlink & $\varHDL$ & $\varHDL$ & $\varHhatDL$\\
		\hline 
	\end{tabular}%
\end{table*}%
In this section, we will compute the ergodic (sum) capacity in the downlink $C_\mathrm{erg}$ and the ergodic (sum) rates when using the ordinary reciprocity relation, instead of the physically consistent one ($R_\mathrm{erg,recip}$) and when the base station ignores the coupling at the base station and at the mobiles ($R_\mathrm{erg,hyp}$). In particular, we compute the (sum) capacity and rate for a given channel and the ergodic ones are obtained by taking the expectation w.r.t. the channel, i.e., for the (sum) capacity
\begin{equation}
C_\mathrm{erg}(P) = \E_{\varHDL} [ C(P) ],
\end{equation}
and in a similar way for the (sum) rates. 
An overview of the different transmit strategies is given in Table~\ref{tab:schemes}.
We assume that the base station obtains an error-free estimate of $\matt D_\mathrm{UL}$ via pilot symbols and that $\vect x~\sim~\mathcal{N}_\mathbb{C}(\vect 0\,\sqrt{\si{\watt}}, \matt R_x)$ with some covariance matrix $\matt R_x$.

\subsection{SU-MISO}
\label{sect:caps}%
For a single user, the channel matrices become vectors. Let
\begin{equation}
\varhDL = \varHDL^T, \quad \vect h_\mathrm{UL}= \matt H_\mathrm{UL}, \quad \vect d_\mathrm{UL} = \matt D_\mathrm{UL}.
\end{equation}

The capacity of the downlink with power $P$ is~\cite{TCAS2_2018}
\begin{equation}
\label{equ:capacity}
C(P) = \log_2 \parens*{1 + \frac{P}{\varsigmathetaDL^2} \norm{ \varhDL}_2^2} \textrm{\ for\ }P_\mathrm{T}=P.
\end{equation}
Let
\begin{equation}
\vect x = \vect f s, \quad s \sim \mathcal{N}_\mathbb{C}(0\,\sqrt{\si{\watt}},P).
\end{equation}
Capacity can be achieved by applying the linear precoder
\begin{equation}
\label{equ:precoderC}
\vect f = \frac{\varhDL^\ast}{\norm{\varhDL}_2}.
\end{equation}
As $\vect f$ can be computed from $\varhDL$, which in turn is computed from $\vect d_\mathrm{UL}$ via \eqref{equ:misoqsimplified}, estimating $\vect d_\mathrm{UL}$ in the uplink and using the physically consistent reciprocity relation \eqref{equ:hdlhulrelation} achieves capacity.

Now consider what happens if the base station uses the information-theoretic model in the up- and downlink, but assumes that the ordinary reciprocity relation $\varHDL = \matt H_\mathrm{UL}^T$ holds, corresponding to $\varhDL = \vect h_\mathrm{UL}$ for SU-MISO. This means it determines the information-theoretic uplink channel $\vect h_\mathrm{UL}$ via \eqref{equ:hul}, and then chooses the optimal precoder based on $\vect h_\mathrm{UL}^\ast$,
\begin{equation}
\vect f_\mathrm{recip} = \frac{\vect h_\mathrm{UL}^\ast}{\norm{\vect h_\mathrm{UL}}_2}
\end{equation}
leading to the rate
\begin{equation}
\label{equ:crecip}
R_\mathrm{recip}(P) = \log_2 \parens*{1 + \frac{P}{\varsigmathetaDL^2} \frac{\abs{\varhDL^H \vect h_\mathrm{UL}^\ast}^2}{\norm{\vect h_\mathrm{UL}}_2^2} } \textrm{\ for\ }P_\mathrm{T}=P.
\end{equation}
Note that this rate is different from \eqref{equ:capacity} and there will be some rate loss compared to capacity.
\newpage
For comparison, let us also consider what happens if the base station ignores the coupling. This means that it does not use Multiport Communication Theory, but rather conventional modeling. To predict how much power the base station radiates, it needs to know the power coupling matrix $\varBhatDL$ that ignores the mutual coupling and uses the mapping
\begin{equation}
\label{equ:transformingcoupling}
\hat{\vect x} = \frac{1}{\sqrt{R_\mathrm{G}}}\varBhatDL^{H/2} \vect u_\mathrm{G}.
\end{equation}
\begin{figure}[!t]
	\centering
	\ifbool{pdffigures}{%
	\includegraphics{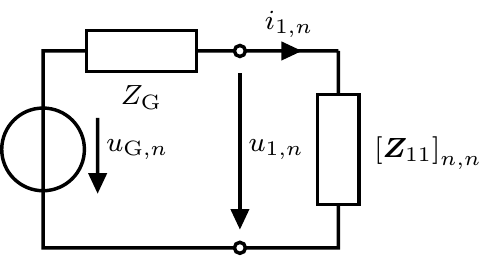}%
	}{%
	\tikzsetnextfilename{simplevoltagedivider}%
	\input{simplevoltagedivider}%
	}%
	\caption{Simplified circuit for measuring $P_{\mathrm{T},n}$.}%
	\label{fig:vdiv}%
\end{figure}%
$\varBhatDL$ is diagonal and its diagonal entries can be obtained by connecting a linear generator to only one antenna in the array at a time, terminating the other antennas with open circuits and measuring the power $P_{\mathrm{T,p},n}$ flowing into the antenna. This means that when the $n$th antenna is excited with the voltage $u_{\mathrm{G},n}$ corresponding to some $\hat{x}_n$, $i_{1,n'} = \SI{0}{\ampere}\ \forall n' \ne n$ and the relevant part of the circuit reduces to a simple voltage divider (Fig.~\ref{fig:vdiv}). The base station predicts that it radiates
\begin{equation}
P_{\mathrm{T,p},n} =  \abs{u_{\mathrm{G},n}}^2 \frac{\Real \parens*{\bracks*{\matt Z_{11}}_{n,n}} }{\abs{\bracks*{\matt Z_{11}}_{n,n}+Z_\mathrm{G}}^2}.
\end{equation}
Similar to \eqref{equ:physicalmodel}, we also have
\begin{equation}
P_{\mathrm{T,p},n} = \abs{u_{\mathrm{G},n}}^2 \frac{\big[ \varBhatDL \big]_{n,n}}{R_\mathrm{G}}.
\end{equation}
Thus analogously to \eqref{equ:defb}, $\varBhatDL$ is given by
\begin{gather}
\begin{split}
\varBhatDL= \aligniftwocolumn R_\mathrm{G} ( \diag(\matt Z_{11}) +  Z_\mathrm{G}\matt I )^{-H} \Real(\diag(\matt Z_{11}))\breakiftwocolumn
            \aligniftwocolumn\cdotiftwocolumn( \diag(\matt Z_{11}) +  Z_\mathrm{G}\matt I )^{-1}
\end{split}
\shortintertext{and}
\varBhatDL^{1/2} = \sqrt{R_\mathrm{G}} ( \diag(\matt Z_{11}) +  Z_\mathrm{G}\matt I )^{-H} \Real(\diag(\matt Z_{11}))^{1/2}.
\end{gather}
If the impedance of all base station antennas is the same, i.e., $\diag(\matt Z_{11})$ is a scaled identity matrix, then $\varBhatDL$ is also a scaled identity matrix. For an arbitrary excitation of the antenna array, the base station predicts the radiated power as~\cite{TCAS2_2018}
\begin{equation}
\label{equ:defPTp}
P_\mathrm{T,p} = \E\big[\norm{\hat{\vect x}}_2^2 \big] =\frac{\E\big[\vect u_\mathrm{G}^H \varBhatDL \vect u_\mathrm{G}\big]}{R_\mathrm{G}}.
\end{equation}
As $\vect x$ is a zero-mean Gaussian random variable and as
\begin{equation}
\label{equ:relxxhat}
\hat{\vect x} = \varBDL^{-H/2} \varBhatDL^{H/2} \vect x,
\end{equation}
$\hat{\vect x}\sim\mathcal{N}_\mathbb{C}(\vect 0\,\sqrt{\si{\watt}}, \matt R_{\hat x})$.
Due to the mapping \eqref{equ:transformingcoupling}, the base station does not transmit over the information-theoretic channel $\varHDL$, but over another information-theoretic channel $\varHhatDL$, the one ignoring the coupling, as given by~\cite{TCAS2_2018}
\begin{equation}
\label{equ:mimoitchannnelmismatch}
\varHhatDL = \varsigmathetaDL \varRetaDL^{-1/2} \varDDL \varBhatDL^{-H/2}.
\end{equation}
For SU-MISO, let $\vect d = \matt D^T$ so that we can define the column vector
\begin{equation}
\varhhatDL^T = \varHhatDL \stackrel{\eqref{equ:noisenorm}}{=} \vect d^T \varBhatDL^{-H/2}.
\end{equation}
When the base station uses a precoder similar to \eqref{equ:precoderC},
\begin{equation}
\label{equ:precoderhyp}
\vect f_\mathrm{hyp} = \frac{\varhhatDL^\ast}{\norm{\varhhatDL}_2},
\end{equation}
it can achieve the (hypothetical) rate~\cite{TCAS2_2018}
\begin{equation}
\label{equ:rhyp}
\begin{split}
R_\mathrm{hyp}(P) &=\log_2 \parens*{1 + \frac{P}{\varsigmathetaDL^2} \norm{ \varhhatDL}_2^2}\textrm{\ for\ }P_\mathrm{T,p}=P.
\end{split}
\end{equation}
We call the rate hypothetical because it is what the base station assumes to achieve. However when base station predicts that it radiates the power $P_\mathrm{T,p}=P$ and uses $\vect f_\mathrm{hyp}$, its (true) radiated power $P_\mathrm{T}\ne P$ in general. Consider the ratio $\alpha$~\cite{TCAS2_2018} that follows from \eqref{equ:physicalmodel}, \eqref{equ:defPTp}, \eqref{equ:relxxhat} and \eqref{equ:precoderhyp},
\begin{equation}
\label{equ:truepower}
\alpha = \frac{P_\mathrm{T}}{P_\mathrm{T,p}}= \frac{\varhhatDL^\ast}{\norm{\varhhatDL}_2} \varBhatDL^{-1/2} \varBDL \varBhatDL^{-H/2} \frac{\varhhatDL^T}{\norm{\varhhatDL}_2}.
\end{equation}
\begin{figure}[!t]
	\centering%
	\ifbool{pdffigures}{%
	\includegraphics{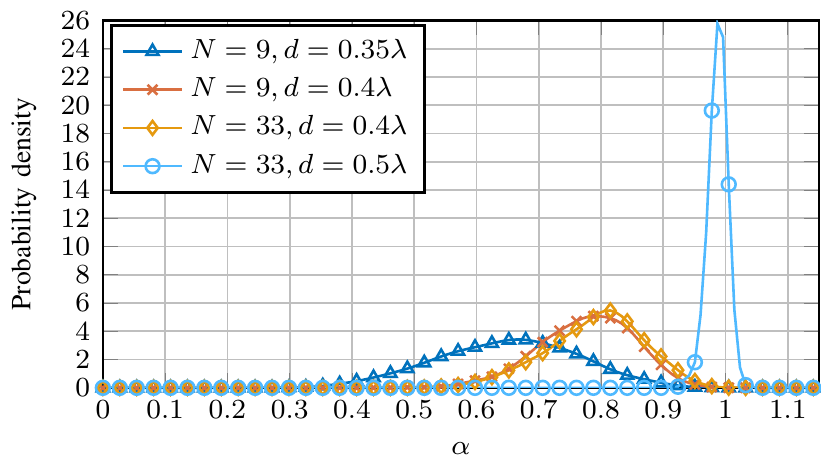}%
	}{%
	\tikzsetnextfilename{compareWSA_alphadlhistUCA}%
	\input{compareWSA_alphadlhistUCA}%
	}%
	\caption{Probability density of $\alpha$ for a uniform circular array (UCA) for four scenarios in a SU-MISO i.i.d. channel.}%
	\label{fig:alphadl}%
\end{figure}%
$\alpha$ is a function of the channel and for some of its realizations, 
\begin{equation}
\alpha < 1 \Leftrightarrow P_\mathrm{T,p} > P_\mathrm{T} \quad \textrm{or} \quad \alpha > 1 \Leftrightarrow P_\mathrm{T,p} < P_\mathrm{T},
\end{equation}
see Figs.~\ref{fig:alphadl} and \ref{fig:qalphadl}, but $P_\mathrm{T,p} < P_\mathrm{T}$ is extremely rare for $d=0.35\lambda$ and $d=0.4\lambda$, but less so for $d=0.5\lambda$. 
Therefore, depending on the channel realization, there will be rate curves that actually require more or less transmit power than predicted. 
On the one hand, if $\alpha > 1$ and if $P_\mathrm{T,p}$ is as large as the power available linearly from the power amplifiers, there will be non-linear distortions due to $P_\mathrm{T} > P_\mathrm{T,p}$, which may cause transmission failure.
On the other hand, if $\alpha < 1$, i.e., $P_\mathrm{T} < P_\mathrm{T,p}$, the transmission will be successful but the power budget is not fully utilized.

The probability densities in this paper are estimated on a grid of 128 points in a Monte Carlo simulation (see Sections~\ref{sect:sim} and \ref{sec:simquadriga}) using the MATLAB implementation~\cite{KDEmatlabcentral} based on the theory in~\cite{KDE} with a Gaussian kernel. Note that $\alpha$ does not need to be estimated in the communication system; it is only introduced to explain the simulation results.

\subsection{SU-MIMO}
For SU-MIMO, the capacity in the downlink with power $P$ is given by~\cite{TelatarCapP2PMIMO}
\begin{equation}
\label{equ:capSUMIMO}
\begin{split}
C(P) = \log_2  \abs*{ \matt I + \varsigmathetaDL^{-2}\varHDL^H \varHDL \matt R_{x}} \textrm{\ for\ }P_\mathrm{T}=P, \breakiftwocolumnquad
\matt R_{x} = \matt V \matt \Psi \matt V^H, \quad \tr(\matt \Psi)=P,
\end{split}
\end{equation}
where $\matt V$ is obtained from the eigenvalue decomposition (EVD)
\begin{equation}
\varHDL^H \varHDL = \matt V \matt \Phi \matt V^H
\end{equation}
and $\matt \Psi$ is a diagonal matrix whose entries are determined via waterfilling. This can be achieved by transmitting $\vect s\sim \mathcal{N}_\mathbb{C}(\vect 0\,\sqrt{\si{\watt}}, \matt \Psi)$ over the precoder $\matt V$, i.e., $\vect x = \matt V \vect s$.

If the base station ignores the mutual coupling at the base station and at the mobiles, it uses \eqref{equ:transformingcoupling} to transform from $\hat{\vect x}$ to $\vect u_\mathrm{G}$ as in SU-MISO, assumes that mobile uses the mapping
\begin{equation}
\label{equ:yhatmapping}
\hat{\vect y} = \frac{\sigma_\vartheta}{\sqrt{R_\mathrm{L}}}\varRetahatDL^{-1/2} \vect u_\mathrm{L}, \quad \varRetahatDL = \diag(\varRetaDL )
\end{equation}
instead of \eqref{equ:consistentyul}, and assumes that
\begin{equation}
\label{equ:bsassumesscaledidentity}
\varthetahatDL \sim \mathcal{N}_\mathbb{C}(\vect 0\,\sqrt{\si{\watt}}, \sigma_\vartheta^2 \matt I)
\end{equation}
holds. Similar to \eqref{equ:capSUMIMO}, the optimal transmit strategy is to choose the precoder $\hat{\matt V}'$ from the singular value decomposition (SVD) of the channel
\begin{equation}
\label{equ:hhatDLdef}
\varHhatDL^\prime = \varsigmathetaDL \varRetahatDL^{-1/2} \varDDL \varBhatDL^{-H/2}= \hat{\matt U}'  \hat{\matt \Phi}^{\prime,1/2} \hat{\matt V}^{\prime,H}
\end{equation}
and the corresponding diagonal power allocation matrix $\hat{\matt \Psi}^\prime$ obtained by waterfilling.

However, the noise distribution at the mobile in the information-theoretic model is
\begin{equation}
\begin{split}
\varthetahatDL \sim \mathcal{N}_\mathbb{C}\parens*{\vect 0\,\sqrt{\si{\watt}}, \varRthetahatDL}, \breakiftwocolumnquad \varRthetahatDL = \varsigmathetaDL^2 \varRetahatDL^{-1/2} \varRetaDL \varRetahatDL^{-H/2},
\end{split}
\end{equation}
contrary to what the base station expects, see \eqref{equ:bsassumesscaledidentity}. Only $\diag\big(\varRthetahatDL\big) = \varsigmathetaDL^2 \matt I$ holds.
This leads to the (hypothetical) rate
\begin{equation}
\begin{split}
R_\mathrm{hyp}(P)=\log_2\abs*{\matt I + \varsigmathetaDL^{-2}\varHhatDL^H \varHhatDL \matt R_{\hat x}} \textrm{\ for\ }P_\mathrm{T,p}=P, \breakiftwocolumnquad
\matt R_{\hat x} = \hat{\matt V}' \hat{\matt \Psi}' \hat{\matt V}^{\prime,H}, \quad \tr(\hat{\matt \Psi}')=P.
\end{split}
\end{equation}
Similar to \eqref{equ:rhyp}, this is only a hypothetical rate, since the true radiated power may be different from the predicted one. By generalization of \eqref{equ:truepower}, we consider the ratio
\begin{equation}
\label{equ:defalphamulti}
\begin{split}
\alpha(P) &\coloneqq \frac{P_\mathrm{T}}{P_\mathrm{T,p}}=\frac{\tr \big(  \varBhatDL^{-1/2} \varBDL \varBhatDL^{-H/2}  \matt R_{\hat x} \big)}{P} = \tr(\matt A(P)),\\
\matt A(P) &=  \varBhatDL^{-1/2} \varBDL \varBhatDL^{-H/2} \hat{\matt V}' \frac{\hat{\matt \Psi}}{P} \hat{\matt V}^{\prime,H}.
\end{split}\raisetag{18pt}
\end{equation}
Contrary to SU-MISO, for SU-MIMO $\alpha$ also depends on the power allocation.

When the base station uses the ordinary reciprocity relation instead of the physically consistent one, the optimal transmit strategy is to use the precoder $\matt V_\mathrm{recip}$ from the EVD
\begin{equation}
\matt H_\mathrm{UL}^\ast \matt H_\mathrm{UL}^T = \matt V_\mathrm{recip} \matt \Phi_\mathrm{recip} \matt V_\mathrm{recip}^H,
\end{equation}
and $\matt \Psi_\mathrm{recip}$ determined via waterfilling. The rate of this scheme is
\begin{equation}
\label{equ:sumimorecip}
\begin{split}
R_\mathrm{recip}(P) \aligniftwocolumn= \log_2 \abs*{\matt I + \varsigmathetaDL^{-2}\matt R_{x} \varHDL^H \varHDL}  \textrm{\ for\ }P_\mathrm{T}=P,\breakiftwocolumnquad
\matt R_{x}\aligniftwocolumn= \matt V_\mathrm{recip} \matt \Psi_\mathrm{recip} \matt V_\mathrm{recip}^H,  \quad \tr(\matt \Psi_\mathrm{recip})=P.\raiseintwocolumn{12pt}
\end{split}
\end{equation}

\subsection{MU-MISO and MU-MIMO}
\label{sect:MU-MISOtheory}
The sum capacity of the MU-MISO/MIMO Broadcast channel (BC) is given by~\cite{ViswanathTseSumCapVectorBC,VishwanathJindalGoldsmith}
\begin{equation}
\label{equ:capmumiso}
C(P) = \max_{\mathclap{\substack{\matt \Xi \succeq \matt 0\,\si{\watt}\\\tr(\matt \Xi)\le P}}} \log_2 \abs*{\matt I + \varsigmathetaDL^{-2} \varHDL^H \matt \Xi \varHDL} \textrm{\ for\ }P_\mathrm{T}=P,
\end{equation}
where $\matt \Xi$ is the (block-)diagonal covariance matrix in the dual Multiple Access Channel (MAC), i.e., it is based on the rate duality between the BC and the dual MAC with the channel $\varHDL^H$. For MU-MIMO, we use the duality from~\cite{HungerJohamRateDuality}, which ensures that streams allocated to the same mobile are orthogonal. Equation \eqref{equ:capmumiso} describes a convex optimization problem that can be solved efficiently by various optimization algorithms, e.g., a projected gradient algorithm~\cite{HungerDissBook} with a step-size control as in~\cite[eq. (14)]{Bertsekas76}. 

For MU-MISO, if the base station ignores the mutual coupling at the base station, it will perform an optimization as in \eqref{equ:capmumiso}, namely
\begin{equation}
\label{equ:capmmmumiso}
R_\mathrm{hyp}(P)= \max_{\mathclap{\substack{\hat{\matt \Xi}\succeq \matt 0\,\si{\watt}\\\tr(\hat{\matt \Xi})\le P}}} \log_2 \abs*{\matt I + \varsigmathetaDL^{-2} \varHhatDL^H \hat{\matt \Xi} \varHhatDL}\textrm{\ for\ }P_\mathrm{T,p}=P.
\end{equation}
Note that this only holds as mutual coupling at the base station does not introduce interference, no matter whether taken into account or not -- as long as the base station has got perfect channel knowledge.

For MU-MIMO, when the base station ignores the mutual coupling at the base station and at the mobiles, or when it uses the ordinary reciprocity relation in the information-theoretic model, the analysis is more involved, since the channel the base station expects and the true channel are different. This is similar to a channel estimation error and this leads to interference. The capacity achieving transmission scheme for perfect channel knowledge is Dirty Paper Coding (DPC). When this scheme is used with a channel estimation error, the achievable rate may even be lower than with linear precoding. This is shown for a lattice-based scheme in a two-user MU-MISO BC in~\cite{DPCCSIErrorBelfiore}. 

When computing the achievable sum rate with linear precoding, a global optimization is required as this problem is non-convex, see e.g., \cite{NikolaGlobalDC,WeberDCOpt}, which optimizes over the transmit covariance in the dual MAC globally. This is only feasible for a small number of users and their antennas. Instead, we use a linear zero forcing (ZF) approach for the comparison that is only guaranteed to find a local optimum. Among several algorithms in the literature~\cite{TSPGuthy2010,ChristensenWSR}, we have chosen LISA~\cite{TSPGuthy2010}, which is an elegant greedy weighted sum rate maximization algorithm with low complexity and very good performance. For the comparison we are considering, the choice of the weighted sum rate maximization algorithm is not substantial.
LISA finds the ZF precoder and power allocation, where we use the variant that does not avoid the matrix inversion to optimize the receive filters. 
Applying it to $\varHDL, \varHhatDL^\prime$ and $\matt H_\mathrm{UL}^T$ for $P_\mathrm{T} = P, P_\mathrm{T,p} = P$ and $P_\mathrm{T} = P$ and transmitting over $\varHDL, \varHhatDL$ and $\varHDL$ respectively, leads to $R_\mathrm{lin}, R_\mathrm{hyp,lin}$ and $R_\mathrm{recip,lin}$. When computing the rates, we do not consider the equalizers at the mobiles, in other words, we assume that they employ an optimum equalization.

\section{Simulations for the I.I.D. Channels}
\label{sect:sim}
In the following, we assume a base station with a uniform circular array (UCA) of $N$ parallel infinitely thin, but perfectly conducting $\lambda/2$-dipoles with antenna spacing $d$, and one or more mobiles with a UCA consisting also of parallel $\lambda/2$-dipoles. Their impedance matrices can be obtained in a similar way as for $\lambda/4$-monopoles as shown in~\cite{IvrlacWSA2016} (which is based on \cite[Ch. 13]{SchelkunoffFriis}), as they are canonical minimum scattering antennas~\cite{MSant,CMSAnt}. Let $Z_\mathrm{A}$ be the self-impedance of the $\lambda/2$-dipoles. We assume the heuristic match $Z_\mathrm{G} = Z_\mathrm{L} = \Real(Z_\mathrm{A})$~\cite{TCAS2_2018}, which matches the real part of the antenna impedance to the purely resistive source and load impedance. $\matt Q^{1/2}$ is obtained by the (lower triangular) Cholesky decomposition of $\matt Q$.

For the noise parameters, we use the measured ones from~\cite[Tables IV \& VI]{LehmeyerJournalPrinted} with a noise bandwidth of $\SI{740}{\kilo\hertz}$, except that we assume $\Real(Z_\mathrm{A})$ as the input impedance of the LNA, so it fits our model. In this section we also assume that the entries of $\matt Z_{21}$ are i.i.d. according to $\mathcal{N}_\mathbb{C}(0\,\si{\ohm},\sigma_z^2)$. In order to obtain reasonable transmit powers, $\sigma_z\nobreak\approx\nobreak\SI{0.019085}{\ohm}$ is chosen, which corresponds to the absolute value of the mutual impedance between two $\lambda/2$-dipoles separated by $1000\lambda$, which is about $\SI{85.7}{\meter}$ at $\SI{3.5}{\giga\hertz}$.

The ergodic (sum) capacity and rates, the average number of active streams and the empirical probability density of $\alpha$ in the following were computed by a Monte Carlo simulation with $1000$ channel realizations.
\subsection{SU-MISO}
\begin{figure}[!t]
	\centering
	\ifbool{pdffigures}{%
	\includegraphics{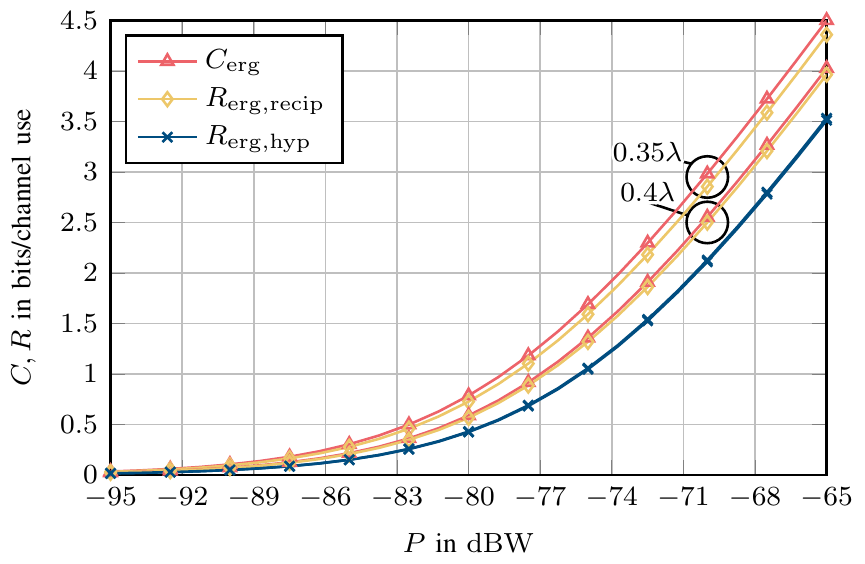}%
	}{%
	\tikzsetnextfilename{compareWSA_DL_UCA_9ant_03}%
	\input{compareWSA_DL_UCA_9ant_0.3}%
	}%
	\caption{Ergodic downlink rates for a UCA with 9 $\lambda/2$-dipoles, and $0.35\lambda$ and $0.4\lambda$ antenna spacing in a SU-MISO i.i.d. channel (based on~\cite{WSAPaper}).}%
	\label{fig:DL803}%
\end{figure}%
\begin{figure}[!t]
	\centering
	\ifbool{pdffigures}{%
	\includegraphics{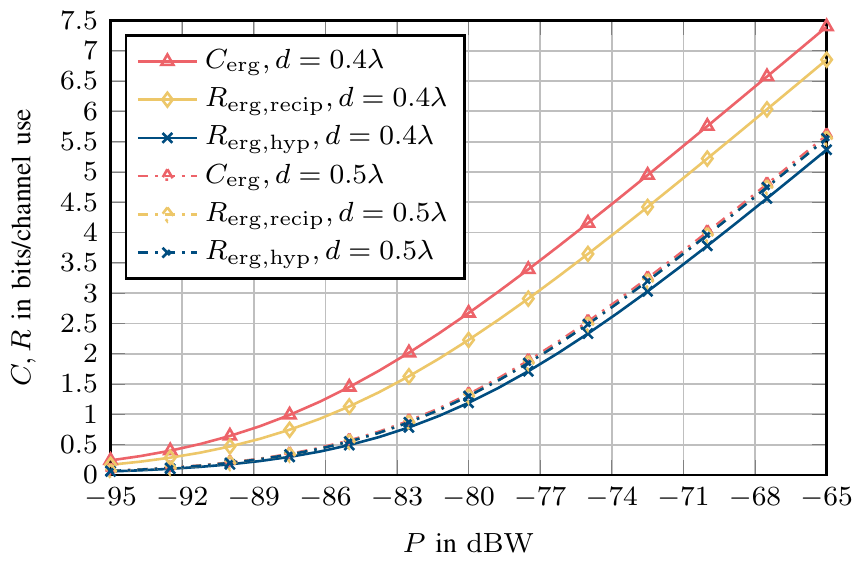}%
	}{%
	\tikzsetnextfilename{compareWSA_DL_UCA_33ant_04}%
	\input{compareWSA_DL_UCA_33ant_0.4}%
	}%
	\caption{Ergodic downlink rates for a UCA with 33 $\lambda/2$-dipoles, and $0.4\lambda$ and $0.5\lambda$ antenna spacing in a SU-MISO i.i.d. channel (based on~\cite{WSAPaper}).}%
	\label{fig:DL3204}%
\end{figure}%
Consider one single antenna receiver in four scenarios: a base station with $N=9$ antennas and $d = 0.35 \lambda$ or $d = 0.4 \lambda$ antenna spacing and one with $N=33$ and $d = 0.4 \lambda$ or $d=0.5\lambda$.  Fig.~\ref{fig:alphadl} shows the probability density for $\alpha$ in these scenarios. The largest variation in $\alpha$ is obtained for a small antenna spacing of $0.35\lambda$, where for some channel realizations only about $\SI{36.2}{\percent}$ of the predicted power is radiated and for some as much as $\SI{94.0}{\percent}$. The variations for $0.4\lambda$ antenna spacing are less pronounced, but there is still a considerable variation in $\alpha$. For $d = 0.5\lambda$ there is even less variation. Furthermore, there is a trend that the larger $d$ is, the further the mass of the distribution of $\alpha$ moves to larger values of $\alpha$. We conclude that the base station radiates on average less power than predicted when it uses conventional modeling. The loss in power is significantly larger for $d=0.35\lambda$ than for $d=0.4\lambda$, and in turn than for $d=0.5\lambda$.

\begin{figure}[t]
	\centering
	\ifbool{pdffigures}{%
		\includegraphics{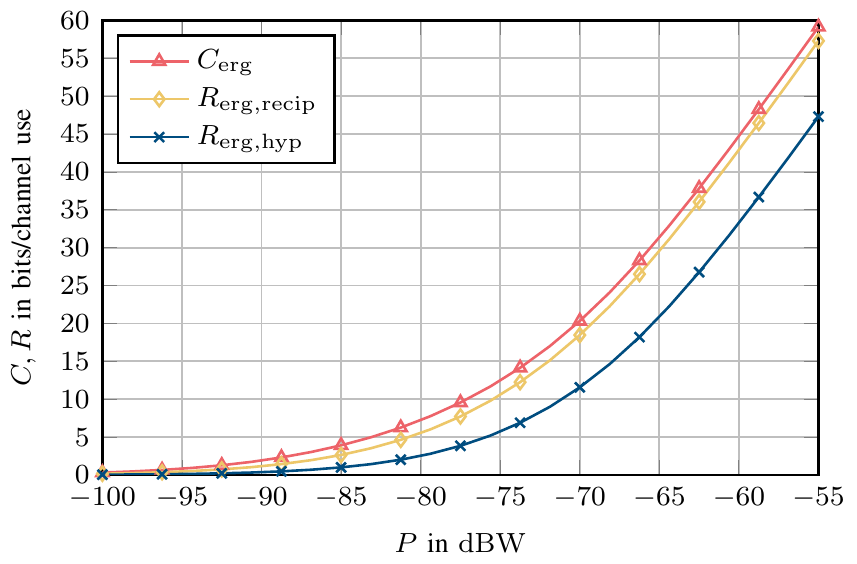}%
	}{%
		\tikzsetnextfilename{compareWSA_MIMODL_UCA_33ant_04}
		\input{compareWSA_MIMODL_UCA_33ant_0.4}%
	}%
	\caption{Ergodic downlink rates for a UCA with 33 $\lambda/2$-dipoles at the base station and a mobile with a UCA with 9 $\lambda/2$-dipoles, both with $0.4\lambda$ antenna spacing, in a SU-MIMO i.i.d. channel.}%
	\label{fig:MIMODL3304}%
\end{figure}%
\begin{figure}[t]
	\centering
	\ifbool{pdffigures}{%
		\includegraphics{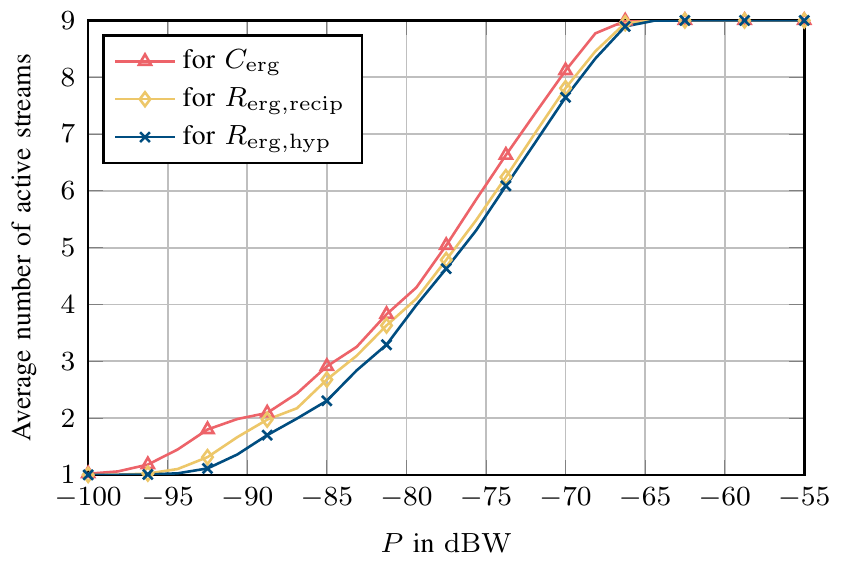}%
	}{%
		\tikzsetnextfilename{compareWSA_nstreamMIMOUCA_33ant}
		\input{compareWSA_nstreamMIMOUCA_33ant}%
	}%
	\caption{Average number of active streams for a UCA with 33 $\lambda/2$-dipoles at the base station and a mobile with a UCA with 9 $\lambda/2$-dipoles, both with $0.4\lambda$ antenna spacing, in a SU-MIMO i.i.d. channel.}%
	\label{fig:MIMODLnstreams}%
\end{figure}%
Figs.~\ref{fig:DL803} and \ref{fig:DL3204} show the ergodic capacities and rates for these scenarios. Comparing them, we can see that for the same $P$, $C_\mathrm{erg}$ and $R_\mathrm{erg,recip}$ are larger for $N=33$ than for $N=9$, and larger for smaller $d$ than for larger $d$. Therefore, a smaller $d$ is advantageous. $R_\mathrm{erg,hyp}$ only changes very little from $d=0.35\lambda$ to $0.4\lambda$, and increases slightly from $d=0.4\lambda$ to $0.5\lambda$. We can also see that using the ordinary reciprocity relation in the information-theoretic model leads to a loss in rate that is small for larger antenna spacings and a small number of antennas, but increases considerably for smaller antenna spacings and a large number of antennas. This loss is caused by the precoder $\vect f_{\mathrm{recip}}$, leading to the beamforming vector $\sqrt{R_\mathrm{G}} \varBDL^{-H/2}  \vect f_{\mathrm{recip}}$. Both are optimal for the ordinary reciprocity relation, but not for the physically consistent one.
Still, using the ordinary reciprocity relation is considerably better than using conventional modeling. $R_\mathrm{erg,hyp}$ shows the same tendency as $R_\mathrm{erg,recip}$, but the gap to $C_\mathrm{erg}$ is significantly larger than for $R_\mathrm{erg,recip}$. This gap is not only caused by a suboptimal precoder, but also by the base station not being able to accurately predict the radiated power $P_\mathrm{T}$ with conventional modeling. Note that mutual coupling is present independent of the antenna separation and does not decrease monotonically with increasing $d$, but rather follows a more complicated relation. It decreases monotonically approximately between $d=0$ and $\lambda/2$, though. If we increase the number of base station antennas further, e.g., to $N=65$, we see that the trends going from $N=9$ to $33$ continue.

\subsection{SU-MIMO}
\begin{figure}[t!]
	\centering
	\ifbool{pdffigures}{%
    \includegraphics{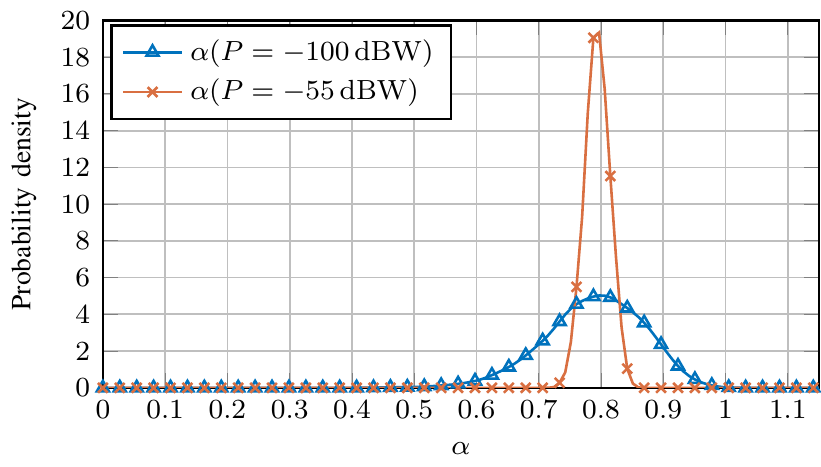}%
    }{%
	\tikzsetnextfilename{compareWSA_alphadlcdfMIMOUCA_33ant_04dst}
	\input{compareWSA_alphadlcdfMIMOUCA_33ant_0.4dst}%
	}%
	\caption{Probability density of $\alpha$ for a UCA with 33 $\lambda/2$-dipoles at the base station and a mobile with a UCA with 9 $\lambda/2$-dipoles, both with $0.4\lambda$ antenna spacing, in a SU-MIMO i.i.d. channel.}%
	\label{fig:MIMODLalphacdf}%
\end{figure}%
Consider a base station with a UCA consisting of $N=33$ $\lambda/2$-dipoles and a mobile with a UCA with $M=9$ $\lambda/2$-dipoles, both with $0.4\lambda$ antenna spacing. Compared to SU-MISO (see Fig.~\ref{fig:DL3204}), the difference between $R_\mathrm{erg,hyp}$ and $C_\mathrm{erg}$ is significantly larger, as shown in Fig.~\ref{fig:MIMODL3304}, although the antenna spacing at the base station is the same. As in SU-MISO, $R_\mathrm{erg,recip}$ achieves a better performance than $R_\mathrm{erg,hyp}$. 

Looking at the average number of active streams (Fig.~\ref{fig:MIMODLnstreams}), all schemes perform similarly. This means the rate difference comes mainly from radiating a different amount of power than predicted and from the suboptimal precoders, instead of a suboptimal number of active streams.

Regarding the predicted radiated power $P_\mathrm{T,p}$ when ignoring the coupling, consider the probability density of $\alpha$ in Fig.~\ref{fig:MIMODLalphacdf}. For $P=\SI{-100}{\decibelwatt}$, the average number of active streams is 1, and the distribution is similar to Fig.~\ref{fig:alphadl}. However for $P=\SI{-55}{\decibelwatt}$, the average number of active streams is close to 9, and the distribution is much more narrow around $\alpha\approx 0.79$. This means that when more streams are active, there is an averaging effect between streams belonging to directions with large $\alpha$ and to those with small $\alpha$. Note that the ratio of the predicted to the radiated power of the individual streams may still experience a distribution similar to when only one stream is active.
\begin{figure}[t]
	\centering
	\ifbool{pdffigures}{%
		\includegraphics{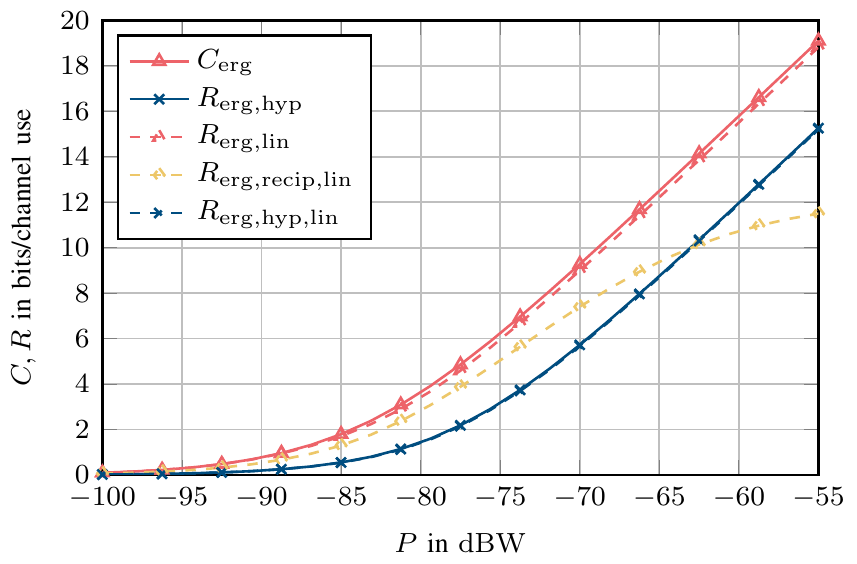}%
	}{%
		\tikzsetnextfilename{compareWSA_MUMISODL_UCA_33ant_04}%
		\input{compareWSA_MUMISODL_UCA_33ant_0.4}%
	}%
	\caption{Ergodic downlink sum rates for a UCA with 33 $\lambda/2$-dipoles with $0.4\lambda$ antenna spacing at the base station and two mobiles in a MU-MISO i.i.d. channel.}%
	\label{fig:MUMISODL3304}%
\end{figure}%
\begin{figure}[t]
	\centering
	\ifbool{pdffigures}{%
		\includegraphics{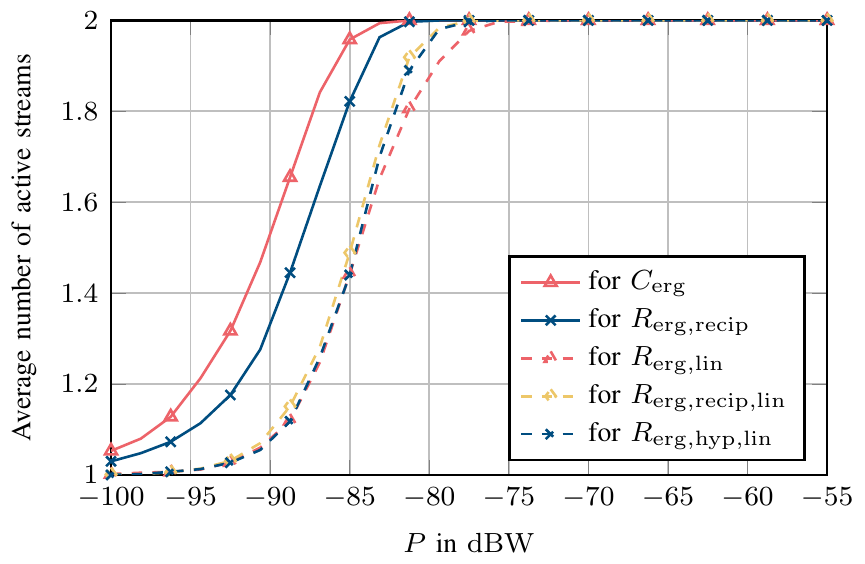}%
	}{%
		\tikzsetnextfilename{compareWSA_nstreamMUMISOUCA_33ant}%
		\input{compareWSA_nstreamMUMISOUCA_33ant}%
	}%
	\caption{Average number of active streams for a UCA with 33 $\lambda/2$-dipoles with $0.4\lambda$ antenna spacing at the base station and two mobiles in a MU-MISO i.i.d. channel.}%
	\label{fig:MUMISODLnstreams}%
\end{figure}%
\subsection{MU-MISO}
Let us compare the ergodic sum rates in Fig.~\ref{fig:MUMISODL3304} for a base station with $N=33$ $\lambda/2$-dipoles in a UCA communicating to two mobiles with one $\lambda/2$-dipole each. The performance of linear ZF precoding is very close to DPC for both, the sum capacity and the hypothetical sum rate, although fewer streams are active with linear precoding up to around $P=\SI{-75}{\decibelwatt}$, see Fig.~\ref{fig:MUMISODLnstreams}. The loss when ignoring the coupling ($R_\mathrm{erg,hyp}$) is qualitatively similar to the loss for SU-MIMO in Fig.~\ref{fig:MIMODL3304}. For $R_\mathrm{erg,recip,lin}$, the loss is smaller than for $R_\mathrm{erg,hyp,lin}$ for small transmit powers -- and as shown in Fig.~\ref{fig:MUMISODLnstreams}, there is only a bit more than one stream active on average. This means that in this region, the system behaves similarly to a SU-MISO system and is mainly noise limited. For larger transmit powers, however, the loss starts to increase significantly when 2 streams are active on average, because using the ordinary reciprocity relation leads to wrong CSI in the downlink and causes interference. The larger this interference is compared to the noise power, the more important it is. For large interference, the system becomes interference limited and $R_\mathrm{erg,recip,lin}$ saturates. This is why starting from $P\approx\SI{-63}{\decibelwatt}$, $R_\mathrm{erg,recip,lin}$ is even worse than $R_\mathrm{erg,hyp,lin}$.

\begin{figure}[t]
	\centering
	\ifbool{pdffigures}{%
    \includegraphics{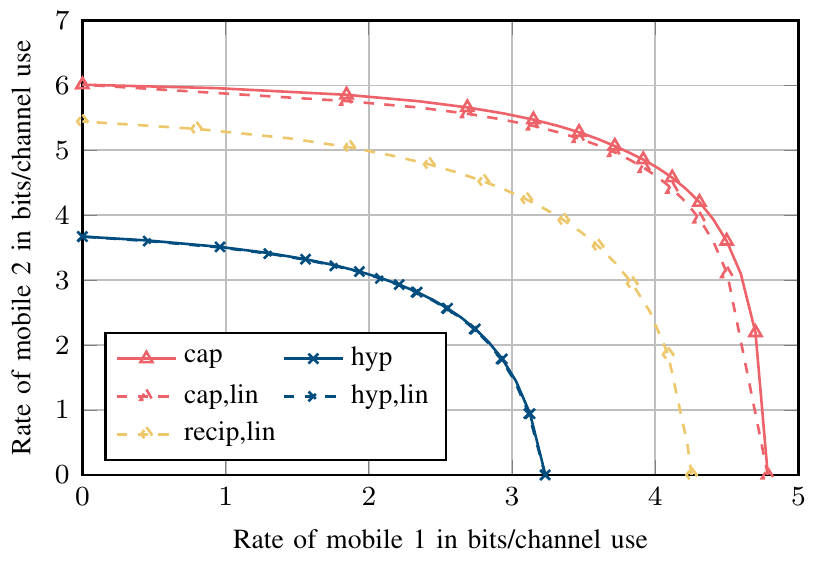}%
    }{%
	\tikzsetnextfilename{rateregionUCA_33ant}%
	\input{rateregionUCA_33ant}
	}%
	\caption{Rate region for one channel realization for a UCA with 33 $\lambda/2$-dipoles with $0.4\lambda$ antenna spacing at the base station and two mobiles in a MU-MISO i.i.d. channel for $P=\SI{-70.57}{\decibelwatt}$.}%
	\label{fig:MUMISOrateregion}%
\end{figure}%
Consider also the rate region for one channel realization. The weighted sum and per-user rates for DPC can be obtained in a similar way to \eqref{equ:capmumiso} and \eqref{equ:capmmmumiso} using, e.g., a projected gradient algorithm~\cite{HungerDissBook}, and for linear precoding using similarly the weighted sum rate maximization algorithm from~\cite{TSPGuthy2010}.
When we have a look at the rate region for one channel realization in the same setting as for the sum rates, we can see a similar behavior as for the sum rate, see Fig.~\ref{fig:MUMISOrateregion}. The curves corresponding to the sum capacity and its corresponding sum rate with linear ZF precoding are called ``cap'' and ``cap,lin'' in the figure, and accordingly ``hyp'' and ``hyp,lin'' for the hypothetical sum rate and ``recip,lin' for the one for the ordinary reciprocity relation. The performance using linear ZF precoding is very close to DPC.

\subsection{MU-MIMO}
\begin{figure}[!t]
	\centering
	\ifbool{pdffigures}{%
		\includegraphics{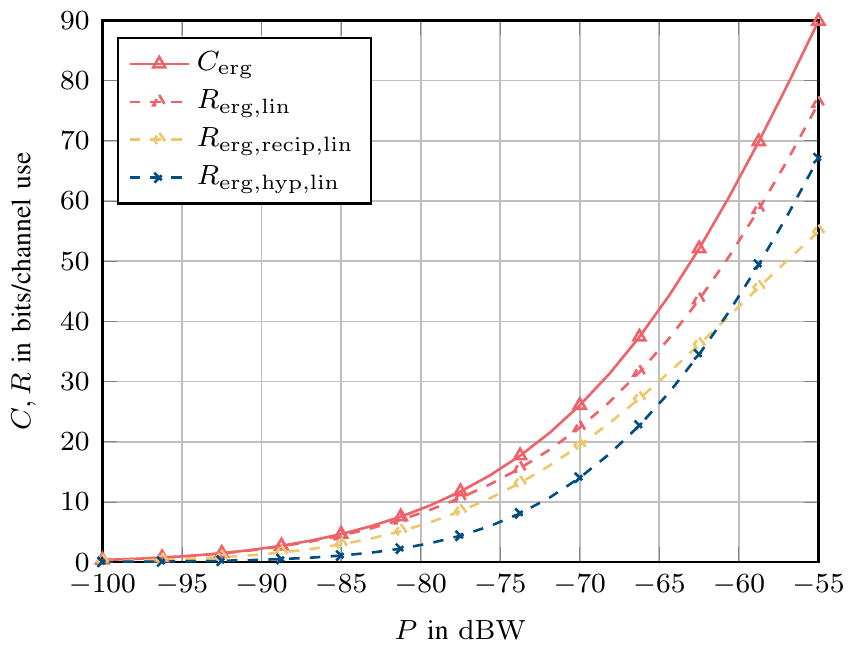}%
	}{%
		\tikzsetnextfilename{compareWSA_MUMIMODL_UCA_33ant_04}%
		\input{compareWSA_MUMIMODL_UCA_33ant_0.4}%
	}%
	\caption{Ergodic downlink sum rates for a UCA with 33 $\lambda/2$-dipoles at the base station and two users with a 9 $\lambda/2$-dipole UCA, all three with $0.4\lambda$ antenna spacing, in a MU-MIMO i.i.d. channel.}%
	\label{fig:MUMIMODL3304}%
\end{figure}%
\begin{figure}[!t]
	\centering
	\ifbool{pdffigures}{%
		\includegraphics{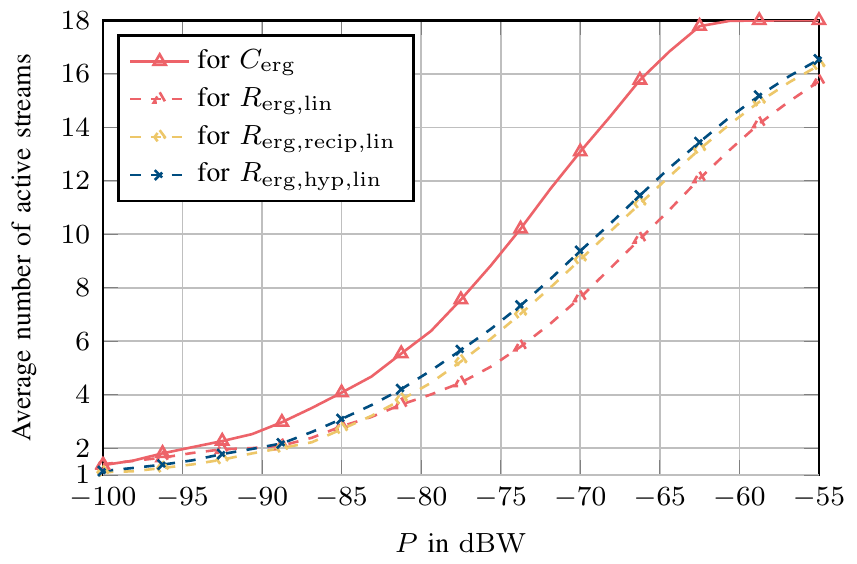}%
	}{%
		\tikzsetnextfilename{compareWSA_nstreamMUMIMOUCA_33ant}%
		\input{compareWSA_nstreamMUMIMOUCA_33ant}%
	}%
	\caption{Average number of active streams for a UCA with 33 $\lambda/2$-dipoles at the base station and two users with a 9 $\lambda/2$-dipole UCA, all three with $0.4\lambda$ antenna spacing, in a MU-MIMO i.i.d. channel.}%
	\label{fig:MUMIMODLnstreams}%
\end{figure}%
For MU-MIMO, let us also consider a UCA with $N=33$ antennas at the base station and two mobiles with a UCA of 9 antennas, all three with $0.4\lambda$ antenna spacing. For smaller transmit powers, the performance of linear ZF precoding is very close to DPC for the sum capacity and the hypothetical sum rate. But as the transmit power increases, the gap also increases, see Fig.~\ref{fig:MUMIMODL3304}. Similar to MU-MISO, for small $P$, $R_\mathrm{erg,recip,lin}$ performs well,  around $P = \SI{-70}{\decibelwatt}$ the gap to $R_\mathrm{erg,hyp,lin}$ starts to decrease considerably, at $P\approx\SI{-61}{\decibelwatt}$ they intersect and for even larger $P$, $R_\mathrm{erg,recip,lin}$ starts to saturate. Compared to MU-MISO, the sum rate loss compared to $C_\mathrm{erg}$ increases for all ergodic rates, i.e., the the loss increases with an increasing number of mobile antennas.

The average numbers of active streams in Fig.~\ref{fig:MUMIMODLnstreams} show that for small transmit powers, they are very similar for DPC and linear ZF, but as the SNR increases, those for DPC increase much faster, i.e., the larger sum capacity and hypothetical rate in Fig.~\ref{fig:MUMIMODL3304} can be explained by DPC supporting more active streams.

\section{Simulations with the QuaDRiGa Channel Generator}
\label{sec:simquadriga}
QuaDRiGa~\cite{QuaDRiGA,QuaDRiGATAP} is a channel generator written in MATLAB, which allows channels to be generated that are more realistic than i.i.d. channels. It is compatible with the current 3GPP channel model, 3GPP TS 38.901~\cite{3GPP38901-1410}, valid from \SI{500}{\mega\hertz} to \SI{100}{\giga\hertz}. As in~\cite{WSA2018Paper}, we consider a single non-sectored base station site in the urban macrocell (UMa) model, but without mobility. The model assumes a hexagonal grid of cells with base station sites at certain corners of the hexagons. When the base station serves all mobiles closest to it, it serves a hexagon with edge length $(500/\sqrt{3})\,\si{\meter}$ and is located at its center. The $\lambda/2$-dipoles at the base station and at the mobiles are all oriented vertically. The mobiles are distributed uniformly in the hexagon outside of a circle with radius $\SI{35}{\meter}$ around the base station. The altitude of the base station is $\SI{25}{\meter}$ and that of the mobiles is determined according to~\cite{3GPP38901-1410}. The continuous time channels obtained by QuaDRiGa need to be scaled by $\Real(Z_\mathrm{A})$ so that they fit the circuit-theoretical model and the receive power matches. Additionally, they need to be filtered by a transmit and a receive filter and sampled, since the QuaDRiGa continuous time channels are impulse trains for each individual channel between a transmit and a receive antenna. As in~\cite{WSA2018Paper}, we use root-raised cosine transmit and receive filters with $\Delta f = \SI{15}{\kilo\hertz}$ and roll-off factor 1 at the center frequency $\SI{3.5}{\giga\hertz}$, because it does not introduce any noise correlations in time-domain after sampling. The bandwidth is similar to an LTE subcarrier. Regarding the noise parameters, the same parameters as for the i.i.d. channels are used, but the noise (co-)variances are scaled by $15/740$, so they match the smaller bandwidth, maintaining the same noise power per bandwidth. We assume that the channel in discrete time is frequency flat, so the channel evaluated at $\nu=\SI{0}{Hz}$ is~\cite{WSA2018Paper}
\begin{equation}
\matt Z_{21} = \sum_{n_\mathrm{path}=1}^{N_\mathrm{path}} \matt Z_{21,n_\mathrm{path}},
\end{equation}
where $N_\mathrm{path}$ is the number of paths of the QuaDRiGa channel and $\matt Z_{21,n_\mathrm{path}}$ are the coefficients corresponding to path $n_\mathrm{path}$.

\begin{figure}[!t]
	\centering
	\ifbool{pdffigures}{%
		\includegraphics{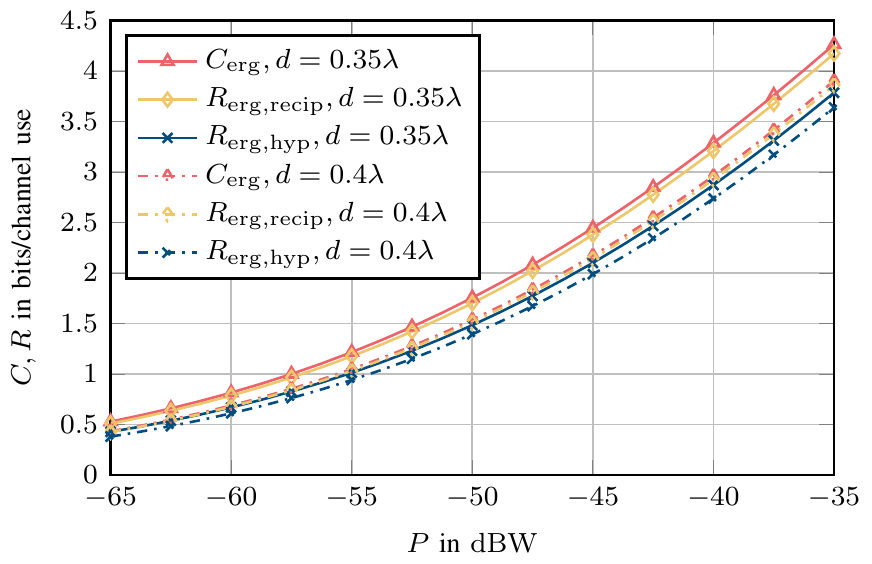}%
	}{%
		\tikzsetnextfilename{qcompareWSA_DL_UCA_9ant_035}%
		\input{qcompareWSA_DL_UCA_9ant_0.35}%
	}%
	\caption{Ergodic downlink rates for a UCA with 9 $\lambda/2$-dipoles, and $0.35\lambda$ and $0.4\lambda$ antenna spacing in a SU-MISO QuaDRiGa channel.}%
	\label{fig:qDL803}%
\end{figure}%
\begin{figure}[!t]
	\centering
	\ifbool{pdffigures}{%
	\includegraphics{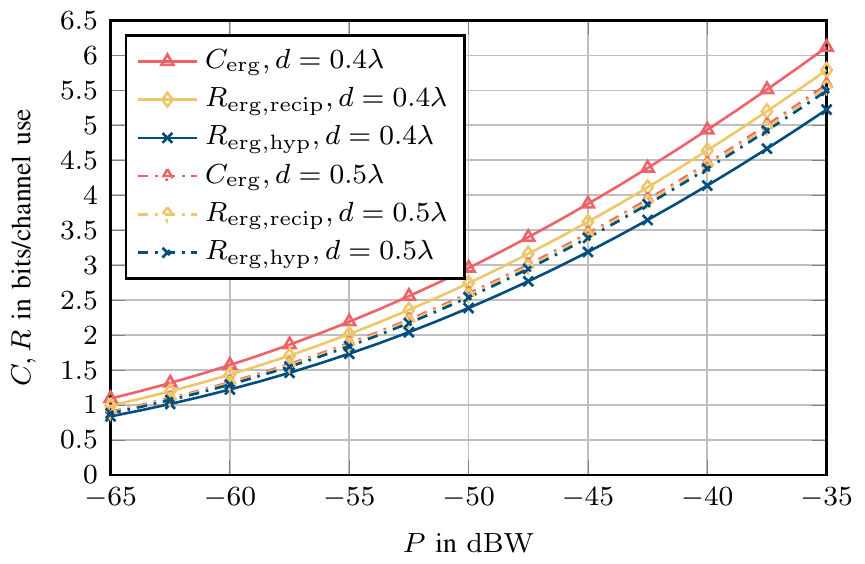}%
	}{%
	\tikzsetnextfilename{qcompareWSA_DL_UCA_33ant_04}%
	\input{qcompareWSA_DL_UCA_33ant_0.4}%
	}%
	\caption{Ergodic downlink rates for a UCA with 33 $\lambda/2$-dipoles and $0.4\lambda$ and $0.5\lambda$ antenna spacing in a SU-MISO QuaDRiGa channel.}%
	\label{fig:qDL3204}%
\end{figure}%
\begin{figure}[t]
	\centering%
	\ifbool{pdffigures}{%
	\includegraphics{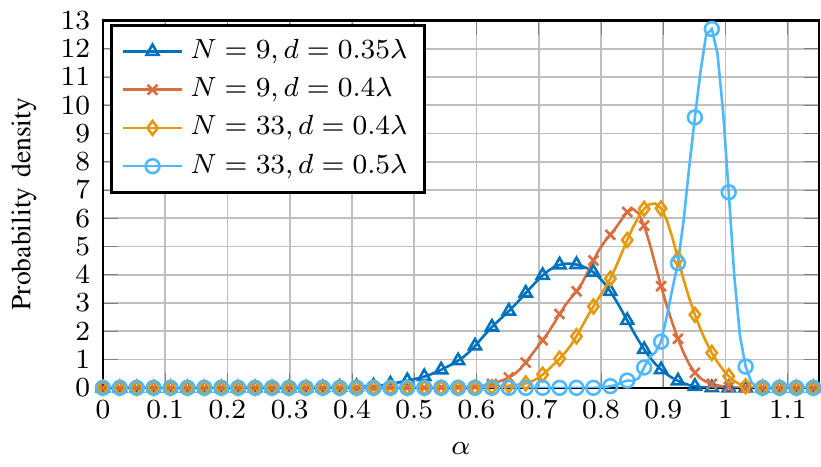}%
	}{%
	\tikzsetnextfilename{qcompareWSA_alphadlhistUCA}%
	\input{qcompareWSA_alphadlhistUCA}%
	}%
	\caption{Probability density of $\alpha$ for a UCA for four scenarios in a SU-MISO QuaDRiGa channel.}%
	\label{fig:qalphadl}%
\end{figure}
\begin{figure}[t]
	\centering
	\ifbool{pdffigures}{%
		\includegraphics{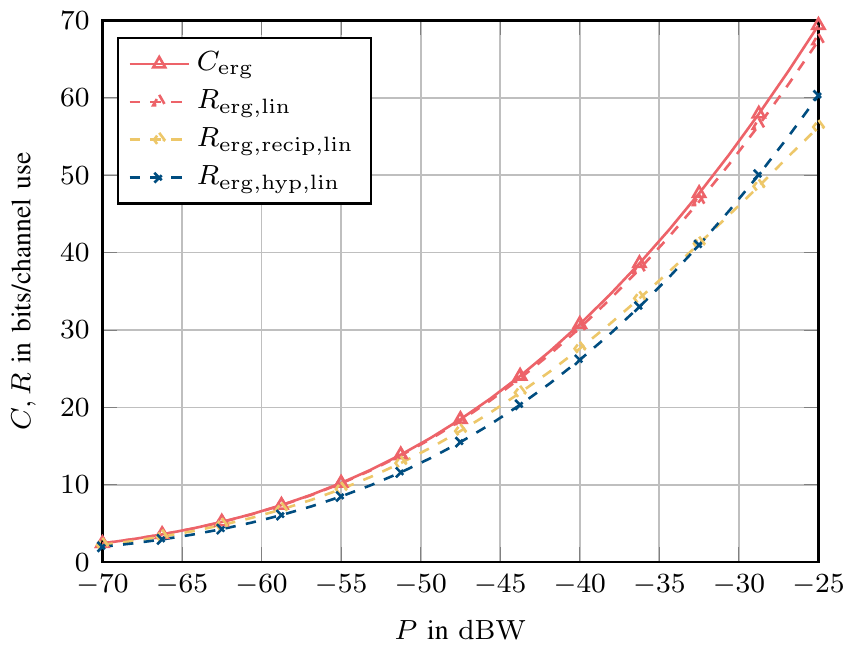}%
	}{%
		\tikzsetnextfilename{qcompareWSA_MUMIMODL_UCA_33ant_04}%
		\input{qcompareWSA_MUMIMODL_UCA_33ant_0.4}%
	}%
	\caption{Ergodic downlink sum rates for a UCA with 33 $\lambda/2$-dipoles at the base station and two users with a 9 $\lambda/2$-dipole UCA, all three with $0.4\lambda$ antenna spacing, in a MU-MIMO QuaDRiGa channel.}%
	\label{fig:qMUMIMODL3304}%
\end{figure}

%
Let us now compare the simulation results for the SU-MISO and MU-MIMO scenarios in the i.i.d. channel with the one in the QuaDRiGa scenario. The attenuation of the channels generated by QuaDRiGa is larger than for the i.i.d. channel, so the ergodic (sum) rates are plotted for a larger $P$ such that similar ergodic (sum) rates are achievable, see Figs.~\ref{fig:qDL803}, \ref{fig:qDL3204} and \ref{fig:qMUMIMODL3304}. For SU-MISO in the range plotted, the slope of the ergodic rates is smaller than in the i.i.d. channel at a similar ergodic rate. This means that for many channel realizations, the channel attenuation is large and the slope of $\log_2(1+\mathrm{SNR})$ is smaller than 1 in logarithmic scale. There is a similar rate loss if the base station uses the ordinary reciprocity relation as in the i.i.d. channel. Similarly, $C_\mathrm{erg}$ and $R_\mathrm{erg,recip}$ are larger for $N=33$ than for $N=9$, and larger for smaller $d$ than for larger $d$. Therefore, a smaller $d$ is also advantageous and desirable here. Also similarly, on average less power is radiated than predicted if the base station uses conventional modeling. For $d=0.35\lambda$ and $0.4\lambda$, the loss due to this  and due to the suboptimal beamforming is smaller for the channels generated by QuaDRiGa, but for $d=0.5\lambda$, they are about the same. This smaller loss for $d=0.35\lambda$ and $0.4\lambda$ corresponds to the distribution of $\alpha$ being shifted a bit closer to 1, see Fig.~\ref{fig:qalphadl}. Furthermore, the variation of $\alpha$ also gets slightly smaller for $d=0.35\lambda$ and $0.4\lambda$, but larger for $d=0.5\lambda$. As in the i.i.d. channel, using the ordinary reciprocity relation leads to higher ergodic rates than conventional modeling.

\begin{figure}[!t]
	\centering
	\ifbool{pdffigures}{%
		\includegraphics{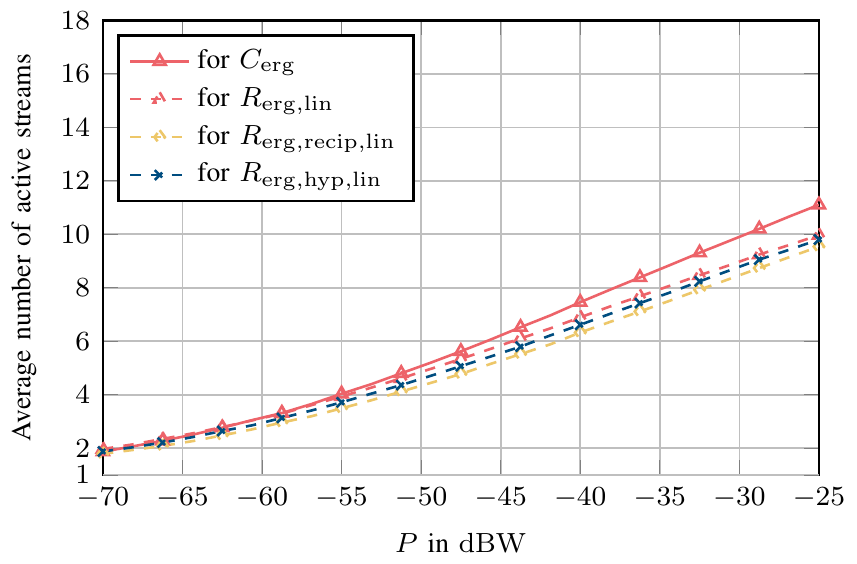}%
	}{%
		\tikzsetnextfilename{qcompareWSA_nstreamMUMIMOUCA_33ant}%
		\input{qcompareWSA_nstreamMUMIMOUCA_33ant}%
	}%
	\caption{Average number of active streams for a UCA with 33 $\lambda/2$-dipoles at the base station and two users with a 9 $\lambda/2$-dipole UCA, all three with $0.4\lambda$ antenna spacing, in a MU-MIMO QuaDRiGa channel.}%
	\label{fig:qMUMIMODLnstreams}%
\end{figure}%
In the MU-MIMO scenario, the ergodic sum rates look similar as in the i.i.d. channels, see Figs.~\ref{fig:MUMIMODL3304} and~\ref{fig:qMUMIMODL3304}, but the losses due to using the ordinary reciprocity relation or ignoring the coupling are smaller and linear ZF is closer to DPC. Regarding the average number of active streams, see Figs.~\ref{fig:MUMIMODLnstreams} and~\ref{fig:qMUMIMODLnstreams}, fewer are active for the same sum rate in the QuaDRiGa channels than in the i.i.d. channels. This can be explained by the larger correlation of the QuaDRiGa per-user channels. Furthermore, the difference between the numbers of active streams for linear ZF and for DPC is smaller for the QuaDRiGa channels. As the linear ZF exploits cooperation between the antennas belonging to the same users using the SVD, the advantage of DPC is that it can remove inter-user interference. Due to the random placing of the users in the QuaDRiGa model, the channels to different users are almost orthogonal, so DPC is less beneficial than for the i.i.d. channel model.

\section{Conclusions}
\label{sec:conclusions}
We have analyzed the reciprocity of a MU-MIMO TDD system based on Multiport Communication Theory. We have seen that by incorporating the physical noise model and the power consistency, the ordinary (pseudo-physical) reciprocity relation between the information-theoretic up- and downlink channel does not hold in general -- even if the noiseless relation between the transmit voltage sources and the receive load voltages is reciprocal. Instead, a physically consistent reciprocity relation holds. We have shown how the base station can achieve capacity using this relation when it computes the downlink channel from the uplink channel: namely, by using the power-coupling matrix it needs to know anyway to obtain the information-theoretic channel and by using a database of noise covariance matrices of the mobiles.

We have shown that when the base station uses the ordinary reciprocity relation, it will use suboptimal beamforming vectors and suboptimal power allocations that can significantly decrease the (sum) rate of the downlink, depending on the array geometry and on the type of antennas used. When the base station uses conventional modeling, i.e., if it ignores the coupling, there can also be a significant rate loss. Furthermore it cannot even predict the power it radiates accurately and the radiated power can vary greatly. In multi-user systems, using the ordinary reciprocity relation is similar to having a channel estimation error, which leads to intra-cell interference between different users. 
The loss in achievable rate when ignoring mutual coupling is larger for a reduced antenna spacing, but capacity increases at the same time. Compactness is therefore advantageous for better performance.

These conclusions hold both for i.i.d. channels and for channels based on the 3GPP TS 38.901 UMa model generated by QuaDRiGa. This highlights the importance of taking the mutual coupling into account, and its effects on the reciprocity in the information-theoretic channel. It is sensible to take it into account by using two matrix multiplications with matrices that can be determined offline at the design stage.

Although our numerical results are based on canonical minimum scattering antennas to enable an analytic calculation of the impedance matrices of the arrays, the analysis is not limited to these types of antenna elements. For other antenna elements, the impedance matrices must be computed numerically with an appropriate electromagnetic solver or must be measured.
Similarly, although we have only provided numerical results for i.i.d. channels and 3GPP TS 38.901 UMa channels, the approach presented is not limited to these types of channels.

\section*{Acknowledgment}
The authors would like to thank C. Mollén for asking inspiring questions motivating them to investigate this topic, and M. T. Ivrlač, who was the main author laying out the Multiport Communication Theory, which is the basis on which this work has been carried out.
The authors would like to acknowledge the contributions of their colleagues in the Horizon 2020 project ONE5G (ICT-760809), although the views expressed in this contribution are those of the authors and do not necessarily represent the project.

\IEEEtriggeratref{31}
\bibliography{transactions1_bibtexnew}

\begin{thebibliography}{10}
\providecommand{\url}[1]{#1}
\csname url@samestyle\endcsname
\providecommand{\newblock}{\relax}
\providecommand{\bibinfo}[2]{#2}
\providecommand{\BIBentrySTDinterwordspacing}{\spaceskip=0pt\relax}
\providecommand{\BIBentryALTinterwordstretchfactor}{4}
\providecommand{\BIBentryALTinterwordspacing}{\spaceskip=\fontdimen2\font plus
\BIBentryALTinterwordstretchfactor\fontdimen3\font minus
  \fontdimen4\font\relax}
\providecommand{\BIBforeignlanguage}[2]{{%
\expandafter\ifx\csname l@#1\endcsname\relax
\typeout{** WARNING: IEEEtran.bst: No hyphenation pattern has been}%
\typeout{** loaded for the language `#1'. Using the pattern for}%
\typeout{** the default language instead.}%
\else
\language=\csname l@#1\endcsname
\fi
#2}}
\providecommand{\BIBdecl}{\relax}
\BIBdecl

\bibitem{LarssonMMimoCommag}
E.~G. Larsson, O.~Edfors, F.~Tufvesson, and T.~L. Marzetta, ``Massive {MIMO}
  for next generation wireless systems,'' \emph{IEEE Commun. Mag.}, vol.~52,
  no.~2, pp. 186--195, Feb. 2014.

\bibitem{PetermannJournal}
M.~Petermann, M.~Stefer, F.~Ludwig, D.~Wübben, M.~Schneider, S.~Paul, and
  K.-D. Kammeyer, ``Multi-user pre-processing in multi-antenna {OFDM} {TDD}
  systems with non-reciprocal transceivers,'' \emph{IEEE Trans. Commun.},
  vol.~61, no.~9, pp. 3781--3793, Sep. 2013.

\bibitem{KaltenbergerMobileSummit2010}
F.~Kaltenberger, H.~Jiang, M.~Guillaud, and R.~Knopp, ``Relative channel
  reciprocity calibration in {MIMO}/{TDD} systems,'' in \emph{Proc. Future
  Netw. Mobile Summit}, Florence, Italy, Jun. 2010.

\bibitem{SamerWenOTACali2016}
S.~Bazzi and W.~Xu, ``A simple over-the-air hardware calibration procedure in
  {TDD} systems,'' in \emph{Proc. IEEE 27th Annu. Int. Symp. Personal Indoor
  Mobile Radio Commun. (PIMRC)}, Valencia, Spain, Sep. 2016, pp. 583--588.

\bibitem{VieiraGlobecom}
J.~Vieira, F.~Rusek, and F.~Tufvesson, ``Reciprocity calibration methods for
  massive {MIMO} based on antenna coupling,'' in \emph{Proc. IEEE Global
  Telecommun. Conf. (Globecom)}, Austin, TX, USA, Dec. 2014, pp. 3708--3712.

\bibitem{VieiraJournal}
J.~Vieira, F.~Rusek, O.~Edfors, S.~Malkowsky, L.~Liu, and F.~Tufvesson,
  ``Reciprocity calibration for massive {MIMO}: {P}roposal, modeling, and
  validation,'' \emph{IEEE Trans. Wireless Commun.}, vol.~16, no.~5, pp.
  3042--3056, May 2017.

\bibitem{Wei}
H.~Wei, D.~Wang, H.~Zhu, J.~Wang, S.~Sun, and X.~You, ``Mutual coupling
  calibration for multiuser massive {MIMO} systems,'' \emph{IEEE Trans.
  Wireless Commun.}, vol.~15, no.~1, pp. 606--619, Jan. 2016.

\bibitem{IvrlacNossekTowardaTheory}
M.~T. Ivrlač and J.~A. Nossek, ``Toward a circuit theory of communication,''
  \emph{IEEE Trans. Circuits Syst. I}, vol.~57, no.~7, pp. 1663--1683, Jul.
  2010.

\bibitem{IvrlacNossekMultiportCommTheory}
------, ``The multiport communication theory,'' \emph{IEEE Circuits Syst.
  Mag.}, vol.~14, no.~3, pp. 27--44, Aug. 2014.

\bibitem{WallaceJensen}
J.~W. Wallace and M.~A. Jensen, ``Mutual coupling in {MIMO} wireless systems: A
  rigorous network theory analysis,'' \emph{IEEE Trans. Wireless Commun.},
  vol.~3, no.~4, pp. 1317--1325, Jul. 2004.

\bibitem{WaldschmidtTVT}
C.~Waldschmidt, S.~Schulteis, and W.~Wiesbeck, ``Complete {RF} system model for
  analysis of compact {MIMO} arrays,'' \emph{{IEEE} Trans. Veh. Technol.},
  vol.~53, no.~3, pp. 579--586, May 2004.

\bibitem{MorrisJensenNetworkModel}
M.~L. Morris and M.~A. Jensen, ``Network model for {MIMO} systems with coupled
  antennas and noisy amplifiers,'' \emph{IEEE Trans. Antennas Propag.},
  vol.~53, no.~1, pp. 545--552, Jan. 2005.

\bibitem{DomiziloiNoise}
C.~P. Domizioli, B.~L. Hughes, K.~G. Gard, and G.~Lazzi, ``Noise correlation in
  compact diversity receivers,'' \emph{IEEE Trans. Commun.}, vol.~58, no.~5,
  pp. 1426--1436, May 2010.

\bibitem{QuaDRiGA}
S.~Jaeckel, L.~Raschkowski, K.~Börner, L.~Thiele, F.~Burkhardt, and
  E.~Eberlein, ``{QuaDRiGa} -- quasi deterministic radio channel generator,
  user manual and documentation,'' Fraunhofer Heinrich Hertz Institute, Tech.
  Rep. Version 2.0.0-664, Aug. 2017.

\bibitem{QuaDRiGATAP}
S.~Jaeckel, L.~Raschkowski, K.~Börner, and L.~Thiele, ``{QuaDRiGa}: A {3-D}
  multi-cell channel model with time evolution for enabling virtual field
  trials,'' \emph{IEEE Trans. Antennas Propag.}, vol.~62, no.~6, pp.
  3242--3256, Jun. 2014.

\bibitem{WSAPaper}
T.~Laas, J.~A. Nossek, S.~Bazzi, and W.~Xu, ``On reciprocity of physically
  consistent {TDD} systems with coupled antennas,'' in \emph{Proc. 21st Int.
  ITG Workshop Smart Antennas (WSA)}, Berlin, Germany, Mar. 2017, pp. 377--382.

\bibitem{SCC2010Ivrlac}
M.~T. Ivrlač and J.~A. Nossek, ``A multiport theory of communications,'' in
  \emph{Proc. 8th Int. ITG Conference Source Channel Coding (SCC)}, Siegen,
  Germany, Jan. 2010.

\bibitem{IvrlacWSA2016}
------, ``On physical limits of massive {MISO} systems,'' in \emph{Proc. 20th
  Int. ITG Workshop Smart Antennas (WSA)}, Munich, Germany, Mar. 2016.

\bibitem{SchelkunoffFriis}
S.~A. Schelkunoff and H.~T. Friis, \emph{Antennas, Theory and Practice}.\hskip
  1em plus 0.5em minus 0.4em\relax Wiley: New York, Chapman \& Hall: London,
  1952.

\bibitem{NoisyFourpoles}
H.~Rothe and W.~Dahlke, ``Theory of noisy fourpoles,'' \emph{Proc. IRE},
  vol.~44, no.~6, pp. 811--818, Jun. 1956.

\bibitem{MIMOWComm}
E.~Biglieri, R.~Calderbank, A.~Constantinides, A.~Goldsmith, A.~Paulraj, and
  H.~V. Poor, \emph{{MIMO} Wireless Communications}, 1st~ed.\hskip 1em plus
  0.5em minus 0.4em\relax Cambridge University Press, 2007.

\bibitem{TCAS2_2018}
T.~Laas, J.~A. Nossek, S.~Bazzi, and W.~Xu, ``On the impact of the mutual
  reactance on the radiated power and on the achievable rates,'' \emph{IEEE
  Trans. Circuits Syst. II, Exp. Briefs}, vol.~65, no.~9, pp. 1179--1183, Sep.
  2018.

\bibitem{KDEmatlabcentral}
\BIBentryALTinterwordspacing
Z.~Botev. (2015, Dec.) Kernel density estimator. Version 1.5.0.0. [Online].
  Available:
  \url{https://www.mathworks.com/matlabcentral/fileexchange/14034-kernel-density-estimator}
\BIBentrySTDinterwordspacing

\bibitem{KDE}
Z.~I. Botev, J.~F. Grotowski, and D.~P. Kroese, ``Kernel density estimation via
  diffusion,'' \emph{Ann. Statist.}, vol.~38, no.~5, pp. 2916--2957, 2010.

\bibitem{TelatarCapP2PMIMO}
E.~Telatar, ``Capacity of multi-antenna {Gaussian} channels,'' \emph{Eur.
  Trans. Telecommun.}, vol.~10, no.~6, pp. 585--595, Nov. 1999.

\bibitem{ViswanathTseSumCapVectorBC}
P.~Viswanath and D.~N.~C. Tse, ``Sum capacity of the vector {Gaussian}
  broadcast channel and uplink--downlink duality,'' \emph{IEEE Trans. Inf.
  Theory}, vol.~49, no.~8, pp. 1912--1921, Aug. 2003.

\bibitem{VishwanathJindalGoldsmith}
S.~Vishwanath, N.~Jindal, and A.~Goldsmith, ``Duality, achievable rates, and
  sum-rate capacity of {Gaussian} {MIMO} broadcast channels,'' \emph{IEEE
  Trans. Inf. Theory}, vol.~49, no.~10, pp. 2658--2668, Oct. 2003.

\bibitem{HungerJohamRateDuality}
R.~Hunger and M.~Joham, ``A general rate duality of the {MIMO} multiple access
  channel and the {MIMO} broadcast channel,'' in \emph{Proc. IEEE Global
  Telecommun. Conf. (Globecom)}, New Orleans, LO, USA, Nov./Dec. 2008.

\bibitem{HungerDissBook}
R.~Hunger, \emph{Analysis and Transceiver Design for the {MIMO} Broadcast
  Channel}, ser. Foundations in Signal Processing, Communications and
  Networking.\hskip 1em plus 0.5em minus 0.4em\relax Springer: Berlin,
  Heidelberg, 2013, no.~8.

\bibitem{Bertsekas76}
D.~P. Bertsekas, ``On the {Goldstein}-{Levitin}-{Polyak} gradient projection
  method,'' \emph{IEEE Trans. Autom. Control}, vol. AC-21, no.~2, pp. 174--184,
  Apr. 1976.

\bibitem{DPCCSIErrorBelfiore}
S.~Yang and J.-C. Belfiore, ``The impact of channel estimation error on the
  {DPC} region of the two-user {Gaussian} broadcast channel,'' in \emph{Proc.
  43rd Annu. Allerton Conf. Commun. Control Comput.}, Monticello, IL, USA, Sep.
  2005, pp. 1366--1372.

\bibitem{NikolaGlobalDC}
K.~Eriksson, S.~Shi, N.~Vucic, M.~Schubert, and E.~G. Larsson, ``Globally
  optimal resource allocation for achieving maximum weighted sum rate,'' in
  \emph{Proc. IEEE Global Telecommun. Conf. (Globecom)}, Miami, FL, USA, Dec.
  2010.

\bibitem{WeberDCOpt}
H.~Al-Shatri and T.~Weber, ``Achieving the maximum sum rate using {D.C.}
  programming in cellular networks,'' \emph{IEEE Trans. Signal Process.},
  vol.~60, no.~3, pp. 1331--1341, Mar. 2012.

\bibitem{TSPGuthy2010}
C.~Guthy, W.~Utschick, R.~Hunger, and M.~Joham, ``Efficient weighted sum rate
  maximization with linear precoding,'' \emph{IEEE Trans. Signal Process.},
  vol.~58, no.~4, pp. 2284--2297, Apr. 2010.

\bibitem{ChristensenWSR}
S.~S. Christensen, R.~Agarwal, E.~de~Carvalho, and J.~M. Cioffi, ``Weighted
  sum-rate maximization using weighted {MMSE} for {MIMO}-{BC} beamforming
  design,'' \emph{IEEE Trans. Wireless Commun.}, vol.~7, no.~12, pp.
  4792--4799, Dec. 2008.

\bibitem{MSant}
W.~K. Kahn and H.~Kurss, ``Minimum-scattering antennas,'' \emph{IEEE Trans.
  Antennas Propag.}, vol.~13, no.~5, pp. 671--675, Sep. 1965.

\bibitem{CMSAnt}
W.~Wasylkiwskyj and W.~K. Kahn, ``Theory of mutual coupling among
  minimum-scattering antennas,'' \emph{IEEE Trans. Antennas Propag.}, vol.~18,
  no.~2, pp. 204--216, Mar. 1970.

\bibitem{LehmeyerJournalPrinted}
B.~Lehmeyer, M.~T. Ivrlač, and J.~A. Nossek, ``{LNA} characterization
  methodologies,'' \emph{Int. J. Circuit Theory Appl.}, vol.~45, no.~9, pp.
  1185--1202, Sep. 2017.

\bibitem{3GPP38901-1410}
``Study on channel model for frequencies from 0.5 to 100 {GHz},'' 3GPP, TS
  38.901, Jun. 2017, {Version} 14.1.0.

\bibitem{WSA2018Paper}
T.~Laas, J.~A. Nossek, S.~Bazzi, and W.~Xu, ``On the impact of the mutual
  impedance of an antenna array on power and achievable rate,'' in \emph{Proc.
  22nd Int. ITG Workshop Smart Antennas (WSA)}, Bochum, Germany, Mar. 2018.

\end{thebibliography}
\end{document}